\documentclass[traditabstract]{aa} 
\usepackage{epsfig}
\usepackage{setspace}
\usepackage{graphicx}
 \usepackage{siunitx}
\usepackage{xcolor}
\usepackage{bm}
\usepackage{amsmath}
\usepackage{txfonts}
\usepackage{amssymb}
\usepackage{natbib}                           
\usepackage{epstopdf}  
\bibpunct{(}{)}{;}{a}{}{,}

\def\compps{{\sc compps}}

\def\Integ{{\em INTEGRAL}}

\def\swift{{\em Swift}} 
\def\maxi{{\em MAXI}} 
\def\Fermi{{\em Fermi}} 
\def\gr{$\gamma$}
\def\chandra{{\em Chandra}} 
\def\xmm{{\em XMM-Newton}} 
\def\igr{IGR~J18245--2452} 
 
\def\be{\begin{equation}} 
\def\ee{\end{equation}} 
 
\begin{document} 

\title{The transitional millisecond pulsar IGR J18245-2452 during its 2013 outburst at X-rays and soft gamma-rays}

\author{V. De Falco\inst{1,2}
\and L. Kuiper\inst{3} 
\and E. Bozzo\inst{4}
\and C. Ferrigno\inst{4}
\and J. Poutanen\inst{5,6} 
\and L. Stella\inst{7}  
\and M. Falanga\inst{2,1}         
} 
 
\offprints{V. De Falco} 

\titlerunning{\Integ\ observations of IGR~J18245--2452}  
\authorrunning{De Falco et al.}  
  
\institute{Departement Physik, Universit\"at Basel, Klingelbergstrasse 82, 4056 Basel, Switzerland\\
\email{vittorio-df@issibern.ch} 
\and International Space Science Institute (ISSI), Hallerstrasse 6, 3012 Bern, Switzerland
\and SRON--Netherlands Institute for Space Research, Sorbonnelaan 2, 3584 CA, Utrecht, The Netherlands 
\and ISDC, Data centre for astrophysics, University of Geneva, Chemin d'\'Ecogia 16, 1290 Versoix, Switzerland
\and Tuorla Observatory, Department of Physics and Astronomy, University of Turku, V\"ais\"al\"antie 20, FI-21500 Piikki\"o, Finland
\and Nordita, KTH Royal Institute of Technology and Stockholm University, Roslagstullsbacken 23, SE-10691 Stockholm, Sweden
\and INAF--Osservatorio Astronomico di Roma, via Frascati 33, 00040 Monteporzio Catone (Roma), Italy  
} 
 
\date{} 
 
\abstract{\igr/PSR J1824--2452I is one of the rare transitional accreting millisecond X-ray pulsars, showing direct evidence of switches between states of rotation-powered radio pulsations and accretion-powered X-ray pulsations, dubbed transitional pulsars. \igr\ with a spin frequency of $\sim254.3$ Hz is the only transitional pulsar so far to have shown a full accretion episode, reaching an X-ray luminosity of $\sim10^{37}$~erg~s$^{-1}$ permitting its discovery with \Integ\ in 2013. In this paper, we report on a detailed analysis of the data collected with the IBIS/ISGRI and the two JEM-X monitors on-board \Integ\ at the time of the 2013 outburst. We make use of some complementary data obtained with the instruments on-board \xmm\ and \swift\ in order to perform the averaged broad-band spectral analysis of the source in the energy range 0.4 -- 250~keV. We have found that this spectrum is the hardest among the accreting millisecond X-ray pulsars. We improved the ephemeris, now valid across its full outburst, and report the detection of pulsed emission up to $\sim60$ keV in both the ISGRI ($10.9 \sigma$) and Fermi/GBM ($5.9 \sigma$) bandpass. The alignment of the ISGRI and Fermi GBM 20 -- 60 keV pulse profiles are consistent at a $\sim25\ \mu$s level. We compared the pulse profiles obtained at soft X-rays with \xmm\ with the soft \gr-ray ones, and derived the pulsed fractions  of the fundamental and first harmonic, as well as the time lag of the fundamental harmonic, up to $150\ \mu$s, as a function of energy. We report on a thermonuclear X-ray burst detected with \Integ, and using the properties of the previously type-I X-ray burst, we show that all these events are powered primarily by helium ignited at a depth of $y_{\rm ign} \approx 2.7\times10^8$ g cm${}^{-2}$. For such a helium burst the estimated recurrence time of $\Delta t_{\rm rec}\approx5.6$ d is in agreement with the observations.} 
 

\maketitle

\section{Introduction}  
\label{sec:intro}
Accreting millisecond X-ray pulsars (AMXPs) are known to be old ($\sim$Gyr) neutron stars (NSs) endowed with relatively weak magnetic fields, $B\approx10^{8-9}$ G \citep[see e.g.][]{Psaltis99,DiSalvo03,Patruno10}. These NSs are hosted in transient low-mass X-ray binaries (LMXBs) that spend most of their time in quiescence and occasionally undergo week- to month-long outbursts. Coherent X-ray pulsations are observed from these systems with frequencies comprised between 180 and 600 Hz and their measured orbital periods range from 40 min to 5 hr 
\citep[see][for reviews on AMXPs]{Poutanen06,patruno12}. 

The AMXP \igr\ was discovered by \Integ\ during observations performed in the direction of the Galactic centre in March 2013 \citep{eckert13}. \swift\ and \chandra\ follow-up observations located the source well within the globular cluster M28 \citep{heinke13, romano13, homan13}, thus providing the first measurement of the source distance at 5.5 kpc \citep[][]{Harris96}. The optical counterpart, confirming the LMXB nature of the system, could be identified by observing large variations in the system magnitude between archival observations during quiescence and follow-up pointings performed shortly after the discovery \citep{Monrad13, Pallanca13a, Cohn13, Pallanca13b}. 

The first thermonuclear burst from the source was caught with \swift/XRT \citep{papitto13a, Linares13} and displayed clear burst oscillations at a frequency of $\sim254.4$ Hz \citep{Patruno13}. A second type-I burst was later reported with \maxi\ \citep{Serino13}. Coherent modulations at a period of $\sim254.33$~Hz were discovered in a dedicated \xmm\ observation campaign, allowing \citet{papitto13c} to also measure the system orbital period ($\sim11.03$~hr) and its projected semimajor axis ($\sim0.76$ lt-s). These properties firmly associate \igr\ with the previously known radio pulsar PSR J1824--2452I in M28 \citep{Manchester05}, thus proving that NSs in LMXBs can switch between accretion powered and rotation powered states. LMXBs discovered to undergo such transitions are named ``transitional millisecond pulsars'' \citep{archibald09,demartino10,demartino14,linares14b,patruno14,bassa14,bogdanov14,bogdanov15}. From now on we refer to the source with the name \igr\, since we focus on its X-ray aspects. Together with other AMXPs like SAX~J1808.4--3658, the first discovered system of this class \citep{Wijnands98}, and IGR~J00291+5934, which displayed the first evidence of a clear spin-up during its outburst \citep{Falanga05}, transitional pulsars represent the most convincing proof of the so-called ``pulsar recycling scenario'' \citep{bk74, Alpar82,r82}. Among the transitional pulsars, \igr\ is the only one that has so far displayed a full X-ray outburst, reaching a peak X-ray luminosity comparable to that of other AMXPs in outburst. Its behaviour in X-rays was shown to be particularly puzzling due to a pronounced variability that has been interpreted in terms of intermittent accretion episodes \citep{ferrigno14}. 

In this work, we concentrate on the 2013 outburst from \igr, carrying out for the first time a detailed spectral and timing analysis of the \Integ\ data. To deepen the study of the accretion event displayed by \igr, soft X-rays data from \xmm\ and \swift\ are also used to better constrain the hard X-ray results obtained with \Integ. We also report, for the first time at millisecond timescales, on the detection of pulsed emission by Fermi/GBM.

\section{Observations and data}
\subsection{INTEGRAL}
\label{sec:integral} 
Our \Integ\ \citep{w03} dataset comprises all the 196 science windows (ScWs), that is, the different satellite pointings each lasting $\sim 2-3$\,ks, performed in the direction of \igr\ from 2013, March 26 at 07:12:00 UTC to April 14 at 23:37:49 UTC. The satellite revolutions involved in the analysis were specifically: 1276 -- 1277, and 1279 -- 1280,
and the dedicated Target of Opportunity (ToO) observation covering the entire revolution 1282. The total effective exposure time on the source was of 216.5~ks. We analysed data from the IBIS/ISGRI coded mask telescope \citep{u03,lebrun03}, covering the 20 -- 300~keV energy band, and from the two JEM-X monitors \citep{lund03}, covering the 3 -- 25~keV energy range. The observation in revolution 1282 was the only one performed in the hexagonal dithering mode, which  allows the target to be constantly kept within the fields of view of both IBIS/ISGRI and JEM-X. For all other revolutions, we retained for the scientific analysis only those ScWs for which the source was located at a maximum off-set angle with respect to the satellite aim point of $<12\fdg0$ for IBIS/ISGRI and $< 2\fdg5$ for JEM-X in order to minimize calibration uncertainties. The reduction of all \Integ\ data was performed using the standard {\sc offline science analysis (OSA)} version 10.2 distributed by the ISDC \citep{c03}. The algorithms for spatial and spectral analysis of the different instruments are described in \citet{gold03}.  

\begin{figure}[h] 
\centerline{\psfig{figure=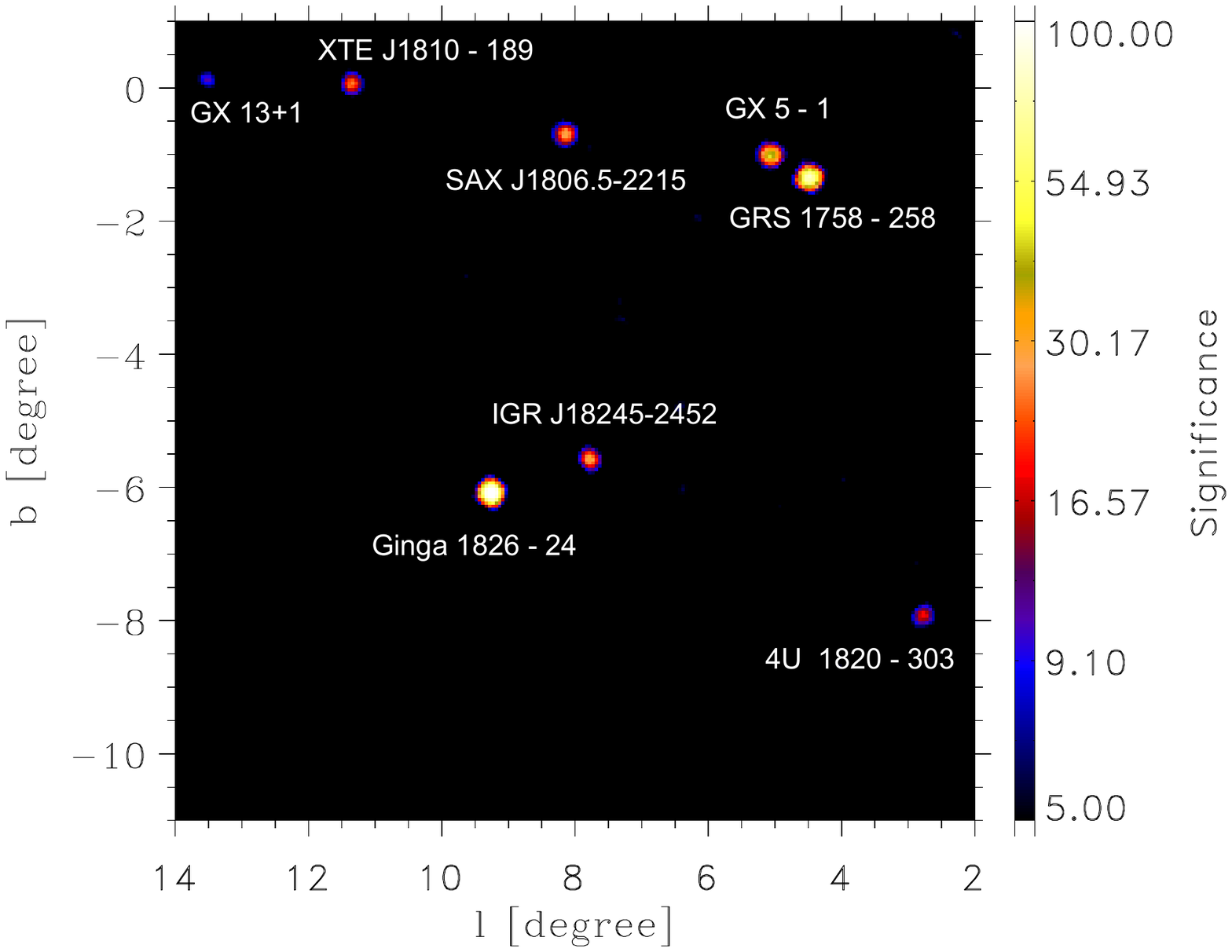, trim=4.5cm 6cm 5cm 8cm, scale=0.48}}
\caption{The \Integ\ IBIS/ISGRI mosaic around \igr\ as obtained in the 20 -- 100 keV energy range 
from the data collected in revolution 1282 (total effective exposure time of $\sim131$~ks). The pixel size in the image corresponds to 3$\arcmin$. \igr\ is detected in the mosaic with a significance of  $\sim 34.5\sigma$.
}
\label{fig:mosa}
\end{figure}

We show in Fig.~\ref{fig:mosa} the ISGRI field of view (significance map) centred on the position of \igr\ as obtained 
from the data in revolution 1282 (20 -- 100 keV energy range). The source is detected in the mosaic with a significance of $\sim 34.5\sigma$. The best determined source position obtained from the mosaic is at 
$\alpha_{\rm J2000} = 18^{\rm h} 24^{\rm m} 33\fs6$ and $\delta_{\rm J2000} = -24{\degr}52\arcmin 48\farcs0$ 
with an associated uncertainty of $0\farcm9$ at the 90\% confidence level (20 -- 100 keV; \citealt{gros03}). We extracted the IBIS/ISGRI light curve of \igr\ with a resolution of one ScW for the entire observational period covered by \Integ\ (see Sect.~\ref{sec:lc}). The JEM-X and ISGRI spectra were extracted using only the data in revolution 1282, as these occurred simultaneously with one of the two available \xmm\ observations (see Sect.~\ref{sec:xmm}) and permitted the most accurate description of the source-averaged broad-band high-energy emission. Three simultaneous \swift/XRT pointings were also available during the same period (see Sect.~\ref{sec:swift}) to complement the \Integ\ and \xmm\ datasets. The averaged broad-band spectrum of the source, as measured simultaneously by all these instruments, is described in Sect.~\ref{sec:spe}. We describe in Sect.~\ref{sec:tm_IGR} the results of the timing analysis of hard X-ray data as obtained from the ISGRI event files. The resulting pulse profiles are compared to the pulse profiles obtained at the soft X-rays with \xmm\ and reported previously by \citet{ferrigno14}.  

\subsection{XMM-Newton}  
\label{sec:xmm}  
\igr\ was observed by \xmm\ \citep{jansen01} twice, on 2013, April 3 -- 4 (obs1, Rev-2439, ID~0701981401) and on 2013, April 13 -- 14 (obs2, Rev-2444, ID~0701981501). The exposure times were 28~ks and 68~ks, respectively. We were interested in the spectra of the source obtained from the RGS (0.4 -- 1.8~keV) and the EPIC-pn (0.5 -- 11~keV) during the observation ID~0701981501 that was carried out during the \Integ\ revolution 1282. We used the average spectra for these instruments (Sect.~\ref{sec:spe}), as published by \citet{ferrigno14}, and refer the readers to this paper for all the details about the data analysis and spectral extraction techniques. We also made use of the EPIC-pn light curve (see Sect.\ref{sec:lc}). We revisited the EPIC-pn timing analysis and  applied the following event selections, ${\rm FLAG}=0$, ${\rm PATTERN}\leq4$ and ${\rm RAWX}=[33,41]$, to the timing data from both observations.  

\subsection{Swift} 
\label{sec:swift}
The \swift/XRT \citep{burrows05} observation campaign carried out to monitor the outburst of \igr\ started $\sim191$ s after the \swift/BAT trigger caused by the source brightening on 2013, March 30 at 02:22:21 UTC \citep{romano13} and lasted until 2013, April 17 at 07:11:51 UTC. The campaign comprises 43 pointings with a total exposure time of $\sim92.7$~ks. We extracted the source light curve of all observations in the 0.5 -- 10~keV energy band (see Sect.~\ref{sec:lc}) by using the XRT online tool \citep{evans09} and performed a detailed analysis of the three pointings that were carried out simultaneously with the \Integ\ revolution 1282. These were pointings: ID~00032785011 performed on 2013, April 13 from 22:01 to 23:50 UTC, ID~00032785012 performed on 2013, April 14 from 06:26 to 09:49 UTC, and ID~00032785013 performed on 2013, April 15 from 17:25 to 19:06 UTC. We processed the \swift/XRT data using standard procedures \citep{burrows05} and the calibration files v.20160113. The considered XRT data were all taken in window-timing (WT) mode and we analysed them by making use of the {\sc xrtpipeline} (v.0.13.2). Filtering and screening criteria were applied by using the FTOOLS contained in the {\sc heasoft}\footnote{http://heasarc.gsfc.nasa.gov/docs/software.html.}software package v.6.19. We extracted source and background light curves and spectra by selecting event grades in the range 0 -- 2. We used the latest spectral redistribution matrices in the HEASARC calibration database. Ancillary response files, accounting for different extraction regions, vignetting and PSF corrections, were generated using the {\sc xrtmarkf} task. The considered data were not found to be affected by any significant pile-up.

\subsection{Fermi/GBM}
\label{instr_gbm}
The Gamma-ray Burst Monitor \citep[GBM;][]{meegan09,bissaldi09} aboard \Fermi\ has as its main goal to increase the science return by observing \gr-ray bursts and other transients below the \Fermi\ LAT \citep{atwood09} passband (20 MeV -- 300 GeV). The GBM comprises a set of 12 sodium iodide (NaI(Tl)) detectors sensitive across the 8 keV to 1 MeV band, and a set of 2 bismuth germanate (BGO) detectors covering the 150 keV to 40 MeV band, and so overlapping with the \Fermi\ LAT passband. The set of non-imaging detectors provides a continuous view on each occulted (by Earth) hemisphere. Since 2012 November 26 (MJD 56257) the GBM in nominal operation mode provides time-tagged events (TTE) with $2\ \mu$s precision, synchronised to GPS every second, in 128 spectral channels, now allowing detailed timing studies at millisecond accuracies. 

\section{Outburst light curve} 
\label{sec:lc} 
\begin{figure}[h] 
\centerline{\epsfig{file=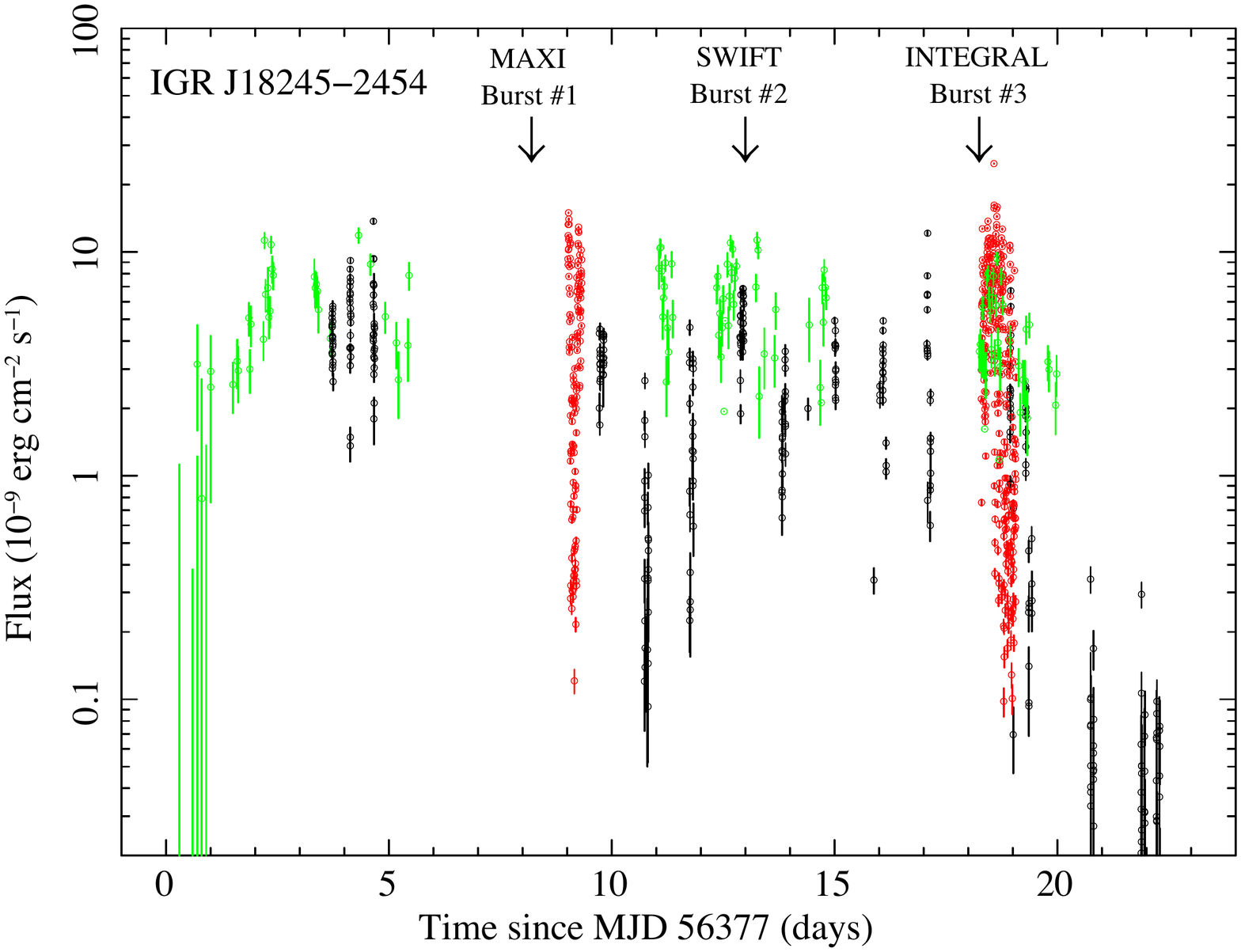, trim=3.5cm 0.5cm 5cm 2.5cm, scale=0.37}} 
\caption{Light curve of the 2013 outburst recorded from \igr\ and observed by \swift/XRT (black circles, bin 
time of 500~s), \xmm/Epic-pn \citep[red circles, with a time bin of 200 s as reported in][]{ferrigno14} and \Integ/ISGRI 
(green circles, the integration time is one science window of $\sim2$ -- 3 ks). The arrows indicate the onset of the 
three type-I X-ray bursts detected from the source (see Sect.~\ref{sec:burst}). \igr\ shows a rapid flux variability.
}
\label{fig:lcr} 
\end{figure}

We report in Fig.~\ref{fig:lcr} the light curve of \igr\ as obtained from all available X-ray data showing that the entire outburst lasts for about 23~days (from 2013 March 26 to April 17). The count-rates measured from all instruments were converted into bolometric flux values (0.4 -- 250 keV) using the spectral analysis results obtained in Sect.~\ref{sec:spe}. 

The global profile of the outburst observed from \igr\ is not too dissimilar from that shown by other AMXPs in outburst, typically characterised by a fast rise time ($\sim 1-2$~d) and a slower decay to quiescence \citep[$\sim4-5$~d; see e.g.][]{Falanga05,bozzo16}. It is the short term variability of the source, observed by all instruments, that is much more peculiar \citep[see Fig.~\ref{fig:lcr} and][]{ferrigno14}. A similar variability is also seen in another two transient AMXPs, PSR J1023+0038 and XSS J1227.0--4859 \citep[see e.g. ][and reference therein]{linares14b}. This rapid flux fluctuation has never been observed in any other AMXP, therefore it constitutes a property remarkably characterising for transient AMXPs. For \igr, this variability has been interpreted as a transition between accretion state and centrifugal inhibition of accretion, possibly causing the launch of outflows \citep{ferrigno14}. However, other models have been proposed to explain the different kind of variability connected to the other two transient AMXPs \citep[see e.g. ][]{DeMartino13,Papitto14,Shahbaz15}. Although \igr\ undergoes dramatic spectral changes on timescales as short as a few seconds, its average spectral energy distribution was identical between the two \xmm\ observations and, more generally, during the entire flat portion of the outburst \citep[from day 3 to 20 in Fig.~\ref{fig:lcr};][]{ferrigno14}. As the fast spectral variations could not be revealed by the reduced sensitivity of the instruments on-board \Integ\ compared to those on-board \xmm, this justified the extraction of a single spectrum for IBIS/ISGRI and the two JEM-X monitors  summing up the data obtained over the entire exposure time available during the revolution 1282. 

During the outburst, three type-I X-ray bursts were recorded from \igr.\ We analyse and discuss the three bursts in Sect.~\ref{sec:burst}.  

\section{Averaged broad-band spectral analysis} 
\label{sec:spe}
In this section, we report on the analysis of the averaged broad-band spectral properties of \igr,\ taking advantage of the hard X-ray energy coverage provided by \Integ.\ We make use of the average JEM-X spectra covering the 5 -- 25 keV energy range and the IBIS/ISGRI spectrum covering the 22 -- 250 keV energy range, as obtained from the data in revolution 1282. We fit these data together with the simultaneous Epic-pn (0.9 -- 11 keV), RGS (0.4 -- 1.8 keV), and XRT spectra (0.4 -- 8 keV). The spectral analysis was carried out using {\sc xspec} version 12.6 \citep{arnaud96}. All uncertainties in the spectral parameters are given at a $1\sigma$ confidential level for a single parameter. All EPIC spectra of the source were optimally rebinned using the prescription in paragraph 5.4 of \citet{Kaastra2016}.

We first fit all the combined spectra by using a simple absorbed power law model, also including a normalisation constant in the fit to take into account uncertainties in the cross-calibrations of the instruments and the source variability (the data are not covering strictly the same time interval, even though they were collected in a compatible portion of the source outburst). This first fit provided an absorption column density of $N_{\rm H}=(0.32\pm0.01) \times 10^{22} {\rm cm}^{-2}$, a power law photon index $\Gamma=1.37\pm 0.01$ and a relatively large $\chi^{2}_{\rm red}{\rm /d.o.f.} = 1.44/2647$, most likely due to the wavy residuals at all energies. Although the fit was not formally acceptable, we noticed that the values of the normalisation constants were all in the range $1.0\pm0.2$, compatible with the finding that the average spectral properties of the source during the considered part of the outburst were stable. A similar range for the normalisation constants was also found in all other fits reported below (we assumed in all cases the EPIC-pn as our reference instrument and fixed its normalisation constant to unity). 

To improve the fit, we modified the power law model by including a cut-off at the higher energies. We measured a cut-off energy of $E_{\rm cut}= 122^{+21}_{-16}$ keV, an absorption column density of $N_{\rm H}=(0.34\pm0.01) \times 10^{22} {\rm cm}^{-2}$ , and a power law photon index of $\Gamma=1.34\pm 0.01$. The improvement in the fit was modest, still resulting in a poorly reduced $\chi^{2}_{\rm red}{\rm /d.o.f.} = 1.38/2647$. 
The addition of a black-body component, possibly related to the thermal emission from (or close to) the NS surface, 
led to a significant reduction of the residuals from the fit and a much more reasonable $\chi^{2}_{\rm red}{\rm /d.o.f.} = 1.16/2640$. The spectral parameters obtained with this model were: $N_{\rm H}=(0.46\pm0.01) \times 10^{22} {\rm cm}^{-2}$, $\Gamma=1.32\pm0.01$, $E_{\rm cut}=94_{-11}^{+14}$ keV, and $kT_{\rm bb}=0.76_{-0.13}^{+0.14}$ keV. Here $kT_{\rm bb}$ represents the temperature of the black-body emission. The black-body radius is $R_{\rm bb}=(6.5\pm0.1)$ m. In all these fits we add a Gaussian line to take into account an iron emission feature around $\sim6.6$~keV, and three Gaussian lines centred at energies of 1.5, 1.8, and 2.2 keV \citep[][]{ferrigno14}.

Following \citet{ferrigno14}, we also tried to fit the averaged broad-band spectrum of \igr\ with a thermal Comptonisation model \citep[{\sc nthcomp}, ][]{Zdziarski96,Zycki99} to take into account the emission produced by a thermal distribution of electrons which Compton up-scatter the soft seed X-ray photons. This model provided a statistically similar good fit as the phenomenological model described above, once a broad iron line peaking at 6.6~keV is included in the fit (the inclusion of the line leads to an improvement of the fit from $\chi^{2}_{\rm red}{\rm /d.o.f.} = 1.26/2643$ to $\chi^{2}_{\rm red}{\rm /d.o.f.} = 1.19/2640$). Although the results of this fit, summarised in Table~\ref{table:spec}, are quantitatively similar to those previously reported by \citet{ferrigno14}, we noticed that our measured absorption column density, $N_{\rm H}=(0.24\pm0.01) \times 10^{22} {\rm cm}^{-2}$, is significantly lower than other values reported in the literature \citep[see also][]{papitto13c,ferrigno14}. We ascribe this difference to the fact that we are using a much broader energy range and also that a thermal component related to the presence of an accretion disk ({\sc diskbb} in {\sc xspec}) was not needed (see below). 

\begin{figure}[h] 
\centering 
\centerline{\epsfig{file=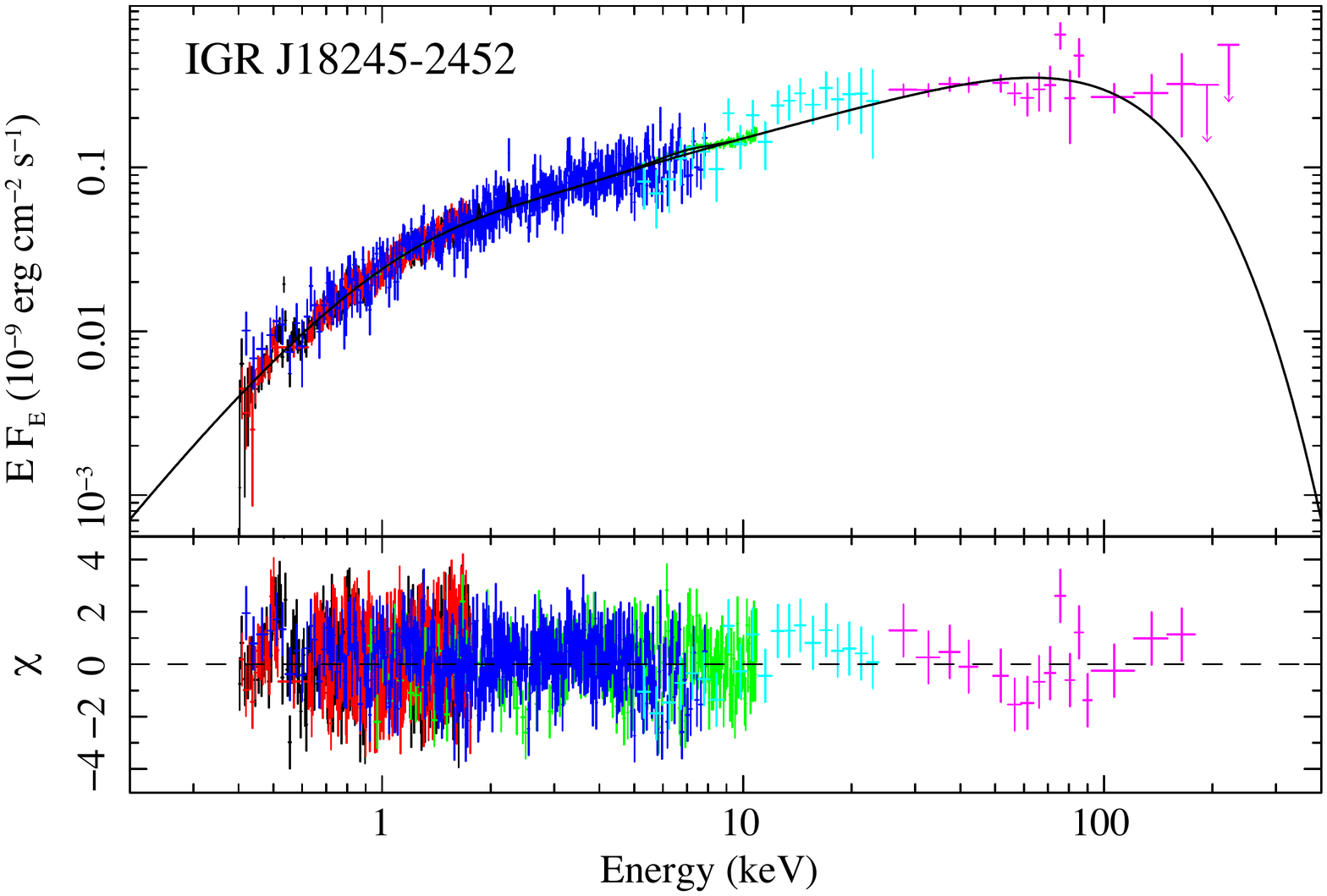, trim=4.5cm 1cm 6cm 4cm, scale=0.38}} 
\caption{Unfolded absorbed broad-band spectrum of \igr\ in the 0.4 -- 250 keV energy range. The data points are obtained from the two \xmm/RGSs (red and black points, 0.4 -- 1.8 keV), \xmm/Epic-pn (blue points, 0.9 -- 11 keV), \swift/XRT (green points, 0.4 -- 8 keV), \Integ/JEM-X (light blue points, 5 -- 25 keV), and \Integ/ISGRI (pink points, 22 -- 250 keV). The fit is obtained with the \compps\ model, represented in the top panel with a solid line. The residuals from the best fit are shown in the bottom panel. This source shows a harder spectrum compared to other AMXPs.}
\label{fig:spe} 
\end{figure}

To compare the averaged broad-band spectrum of \igr\ with those of other AMXPs observed at hard X-rays with \Integ\ \citep[e.g.][]{gdb02,gp05,Falanga05,mfb05,mfc07,ip09,falanga11,falanga12}, we also performed a spectral fit using a thermal Comptonisation model in the slab geometry \citep[\compps, ][]{ps96}. The main parameters are the absorption column density $N_{\rm H}$, the Thomson optical depth $\tau_{\rm T}$ across the slab, the electron temperature $kT_{\rm e}$, the temperature $kT_{\rm bb}$ of the soft-seed thermal photons (assumed to be injected from the bottom of the slab), and the inclination angle $\theta$ between the slab normal and the line of sight. The results of this fit are also reported in Table~\ref{table:spec} and are similar to those measured from other AMXPs in outburst, but the optical depth was $50\%$ larger compared with, for example, XTE J1751--305, which has nearly identical electron temperature \citep{gp05}. We note that the broad Gaussian iron line at $\sim$6.6~keV was also required in this model. The absorption column density measured from the fit with the \compps\ model is compatible with that obtained before using the {\sc nthcomp} model. Our $N_{\rm H}$ value is close to the Galactic value, $0.18\times10^{22}$ cm$^{-2}$, reported in the radio maps of \citet{dickey90,kalberla05}. We note, that combining the different \xmm\ observations with the other spectra, we could not find any evidence of the {\sc diskbb} component from the residuals of our best fit Comptonisation models (see also Fig.~\ref{fig:spe}). The temperature of the {\sc diskbb} was $kT_{\rm diskbb} = (2\pm1)\times10^{-2}$ keV, two orders of magnitude lower than the value reported by \citep{ferrigno14}. We noticed that using the {\sc tbnew} photoelectric absorption model with variable abundances of iron and oxygen \citep{wilms00}, the {\sc diskbb} becomes significant in the fit. This may be due to a decrease of the iron abundances and so the {\sc diskbb} component fits  the data well \citep{ferrigno14}.

\begin{table}[h] 
{\small
\caption{\label{table:spec} Optimal spectral parameters determined from the fits to the average broad-band spectrum 
of \igr\ performed with the {\sc nthcomp} and \compps\ models. The Gaussian lines at 1.5, 1.8, 2.2 keV, and a broad Gaussian iron line were also included in the fit \citep[see][for further details]{ferrigno14}. In both models the {\sc diskbb} component does not improve the fit.}
\centering
\begin{tabular}{lccc} 
\hline 
& \sc{nthcomp} & \compps \\
\hline 
\noalign{\smallskip}  
$N_{\rm H}\ (10^{22} {\rm cm}^{-2})$ & $0.24\pm0.01$ & $0.23\pm0.01$\\ 
$kT_{\rm bb}$ (keV)& $0.34\pm0.01$ & --\\ 
$\Gamma$ & $1.44\pm0.01$ & --\\
$kT_{\rm e}$ (keV)& $23\pm2$ & $30\pm3$\\ 
$kT_{\rm seed}$ (keV)& -- & $0.37\pm0.01$\\ 
$\tau_{\rm T}$ & -- & $2.7^{+1.0}_{-0.1}$\\ 
$\cos \theta $ & -- & $0.76\pm0.02$\\
$A_{\rm seed}\ ({\rm km}^2)$ & -- & $250\pm40$\\
$E_{\rm Fe}$ (keV) & $6.6\pm0.2$ & $6.5\pm0.1$\\
$\sigma_{E_{\rm Fe}}$ (keV) & $1.11\pm0.12$ & $1.09\pm0.13$\\
$\chi^{2}_{\rm red}/{\rm dof}$ & 1.19/2642 & 1.15/2640 \\
$F_{\rm bol}$ ($10^{-10}$ erg cm$^{-2}$ s$^{-1}$)\tablefootmark{a} & $4.25\pm0.02$ & $4.19\pm0.02$\\
\noalign{\smallskip}  
\hline  
\end{tabular}  
\tablefoot{ \tablefoottext{a}{Unabsorbed flux in the 0.4 -- 250 keV energy range.}}
\label{tab:spe} 
}
\end{table} 

We verified that the relatively large $\chi^2_{\rm red}$ (1.15 -- 1.19) in both fits is mainly due to statistical fluctuations of the data and the high statistics available. We could not find any systematic trends in the residuals that could suggest the presence of additional spectral components or that the selected models are not adequate for the fit. The improvement of the \compps\ compared to the {\sc nthcomp}, as statistically evaluated with the F-test, is highly significant, since we obtained an F-test probability of $2.5\times10^{-20}$. An advantage of the \compps\ model is that it allows us to estimate the apparent area of the thermally emitting region on the NS surface, $A_{\rm seed}\approx 250\ (d/5.5\ \mbox{kpc})$ km$^2$. At the distance of \igr\ ($d=5.5$~kpc), the radius of this region is $\approx 8$~km, assuming a canonical NS radius of 10 km and a spherical geometry. This source shows a harder spectrum, with a larger emitting area, and a smaller seed temperature compared to the other AMXPs \citep[e.g.][]{gdb02,gp05,Falanga05,mfb05,mfc07,ip09,falanga11,falanga12}. The spectral slopes for AMXPs have been found in the range $\Gamma\approx(1.8-2.0)$, that is, harder spectra compared to the photon index of $\Gamma\approx1.3$ for \igr.
We show in Fig.~\ref{fig:spe} the absorbed unfolded averaged broad-band spectrum of \igr,\ together with the residuals from the best fit model. 

\section{Timing analysis}
\label{sec:tm_IGR}

\subsection{The pulsar ephemeris}
\label{subsec:pe}
We studied the pulse profile of \igr\ at soft X-rays (0.5 -- 10 keV) and hard X-rays/ soft \gr-rays (20 -- 150 keV) using
\xmm/Epic-pn, and \Integ/ISGRI and \Fermi/GBM data, respectively. Irrespective of the instrument involved, the timing analysis starts with the conversion of the arrival times of the (selected) event registered at the satellite to the solar system barycentre. This process uses the instantaneous spacecraft ephemeris (position and velocity) information, the JPL solar system ephemeris information (DE200 or DE405, we used DE200) and an accurate source position to convert the recorded satellite times from Terrestial Time scale (TT or TDT, which differs from Coordinated Universal Time (UTC) by a number of leap seconds plus a fixed offset of 32.184 s) into Barycentric Dynamical Time (TDB) scale, a time standard for Solar system ephemerides. 

We used the \igr\ source position as listed in Table 1 of \cite{papitto13c}. This position is consistent at (sub)arcsecond level with the most accurate locations reported at optical wavelengths \citep[see e.g.][]{Pallanca13a} and at radio frequencies \citep[see e.g.][]{Pavan13}, and from earlier \chandra\ X-ray observations of M28 \citep{Becker03}. Subsequently, we corrected the TDB arrival times for the acceleration effects along the binary orbit adopting the orbital parameters from Table 1 of \cite{papitto13c}. We analysed the \xmm/Epic-pn data taken in timing mode (timing accuracy $\sim 30\ \mu$s) from both \xmm\ observations (obs. ids. 0701981401 and 0701981501, which are separated in time by $\sim 9.5$ days; see Section \ref{sec:xmm}) performed during the April 2013 outburst. Using the spin frequency of
$254.333\,031\,01(62)$ Hz, as derived from the spin period value and its uncertainty as listed in Table 1 of \cite{papitto13c}, yielded highly significant pulse-phase distributions. However, we noticed a considerable misalignment of $\sim 0.15$ in phase of both pulse-pulse distributions, which is too large when timing data are combined from observations covering periods of weeks, as is the case for \Integ/ISGRI and \Fermi/GBM. Such a shift is indicative of a slightly incorrect spin frequency, or, less likely, is related to the use of the DE200 solar system ephemeris in the barycentering process in our analysis, while the \igr\ parameters of Table 1 of \cite{papitto13c} have been derived adopting DE405. 

Irrespective of the cause of the misalignment, we revisited the `best' spin-frequency, now adopting DE200, because in the barycentering process of the \Integ/ISGRI and \Fermi/GBM timing data we will  also adopt the DE200 solar system ephemeris. We derived through $Z_1^2$ test-statistics \citep{buccheri1983} optimisation using the combined \xmm\ datasets a slightly different spin frequency $\nu$ of $254.333\,030\,87(1)$ Hz. The performed timing analysis improves the spin frequency and confirms the other values reported by \citet{papitto13c}. This procedure also ensures that the pulse phases of the events from both \xmm\ datasets are phase connected automatically across the datagap of $\sim$ 9.5 days. This is demonstrated in Fig. 4 panels a -- c and d -- f showing that the pulse profiles, folded upon one single spin and orbital ephemeris, are nicely aligned. The quoted uncertainty is the statistical error at $3\sigma$ confidential level; the systematical uncertainty in the spin-frequency due to the positional uncertainty \citep[see e.g.][for the method]{Sanna17} in the coordinates of \igr\ is about $6\times 10^{-8}$ Hz for an assumed uncertainty in source location of $0\farcs5$, and thus considerably larger than the statistical one. 
To assess the effects on phase alignment by using a different solar system ephemeris, DE405, as adopted in \cite{papitto13c}, we repeated the frequency optimisation, and obtained an optimum spin frequency value that differs only $+6 \times 10^{-9}$ Hz from the DE200 value. We, therefore, excluded a different solar system ephemeris, DE405, as the cause of the phase misalignment. 

\begin{figure}[h] 
\centering
\epsfig{figure=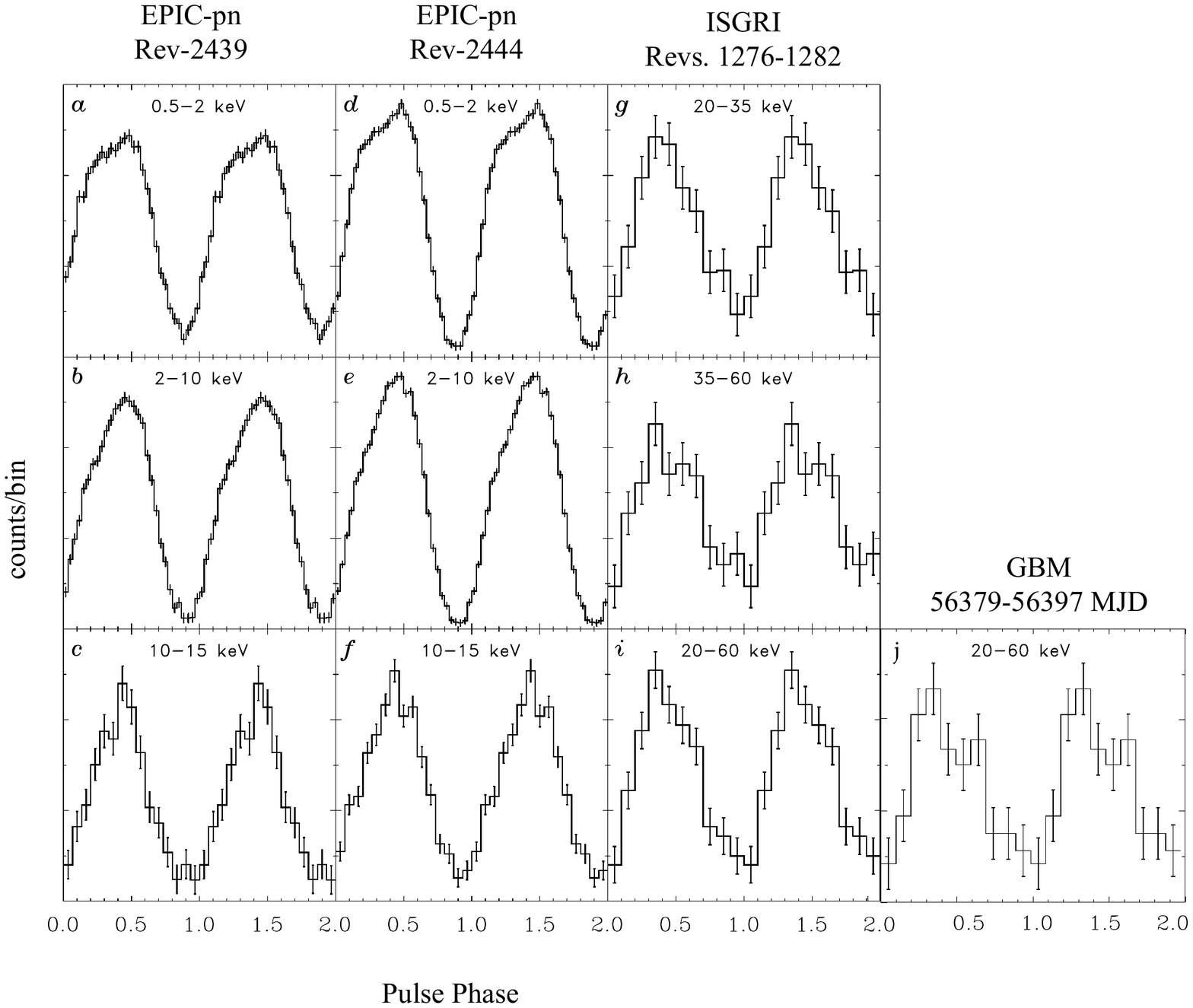, trim=1cm 9cm 1cm 3cm, scale=0.43}
\caption{Pulse profile compilation of \igr, showing the X-ray profiles for \xmm/Epic-pn (panel a -- f), \Integ/ISGRI (panel g -- i), and \Fermi/GBM (panel j). The shift between the 20 -- 60 keV ISGRI and GBM profiles is only $23\pm109\ \mu$s indicating a consistent alignment. This consistency is obtained by an improvement of the spin frequency value.}
\label{fig:tim}
\end{figure}

Application of the newly derived spin-frequency value, which is $1.6\times 10^{-7}$ Hz smaller than the value of \citet{papitto13c}, in the folding procedure now yielded a consistent alignment between the 2 -- 10 keV \xmm\ pulse-phase distributions of the two different \xmm\ observations (see also Fig.~\ref{fig:tim} panels b and e).

\subsection{Pulse profile, pulsed fraction, and time lag}
With an accurate ephemeris, that keeps phase alignment across at least ten days, we proceeded with the timing analysis of ISGRI soft \gr-ray data. We selected \Integ\ observations from the late stage of Rev. 1276 (beyond and including Scw 830010), 1277, 1279 -- 1280 and (ToO) 1282 with \igr\ at $\leq14\fdg5$ from the pointing
axis. We excluded events recorded during time periods where the ISGRI count-rate behaved erratically (e.g. near perigeum ingress/egress, or during periods corresponding to a high solar activity). Additionally, we selected only those events with rise times between channels 7 and 90 \citep{lebrun03} from non-noisy pixels having a pixel illumination 
factor at least $25$\%. 

The outburst averaged 20 -- 60 keV pulse-phase distribution deviates from uniformity at a $10.9\sigma$, applying a $Z_1^2$-test. For the 20 -- 35 keV and 35 -- 60 keV bands, separately, we found significances of $8.5\sigma$ and $6.7\sigma$, respectively, while a hint ($\sim 2\sigma$) was seen in the 60 -- 150 keV band. The ISGRI pulse-phase distributions are shown in the right panels of Fig.~\ref{fig:tim} labelled g, h, i. In the same figure, the (time-averaged) phase-distributions are shown from both \xmm\ observations for the 0.5 -- 2 keV (top; a and d), 2 -- 10 keV (middle; b and e) and 10 -- 15 keV (bottom; c and f) energy bands. We would like to point out that the morphology changes as a function of energy for these X-ray profiles.

We have also folded the barycentered time stamps (accuracy $2\ \mu$s; TTE mode) of the NaI detectors of the \Fermi/GBM, collected during 56379 -- 56397 MJD (2013, March 28 -- April 15; continuously monitoring) using our updated \igr\ DE200 ephemeris. Because of the non-imaging nature of these detectors, we have screened the data only by making selections on observational conditions such as on pointing direction, Earth zenith angle, and spacecraft location with respect to the South Atlantic Anomaly. We also ignored episodes of (intense) bursts. The averaged exposure per NaI detector was 201.3 ks.

For the 20 -- 60 keV band we detected pulsed emission at a $5.9\sigma$ confidential level with a pulse shape fully consistent with the ISGRI 20 -- 60 keV profile, while below 20 keV and above 60 keV we found significances of $3.3\sigma$ and $1.7\sigma$, respectively. The \Fermi/GBM 20 -- 60 keV pulse-phase distribution is shown in Fig.~\ref{fig:tim} (panel j) along with the 20 -- 60 keV ISGRI profile (panel i). Cross-correlation of both profiles (both in 60 bins) shows that the alignment between ISGRI and GBM profiles is fully consistent; we found an insignificant shift of only $23\pm 109\ \mu$s (i.e. $0.006 \pm 0.03$ in phase) between both detectors, validating our updated DE200 ephemeris of \igr.

\begin{figure}[h] 
\centering
\epsfig{figure=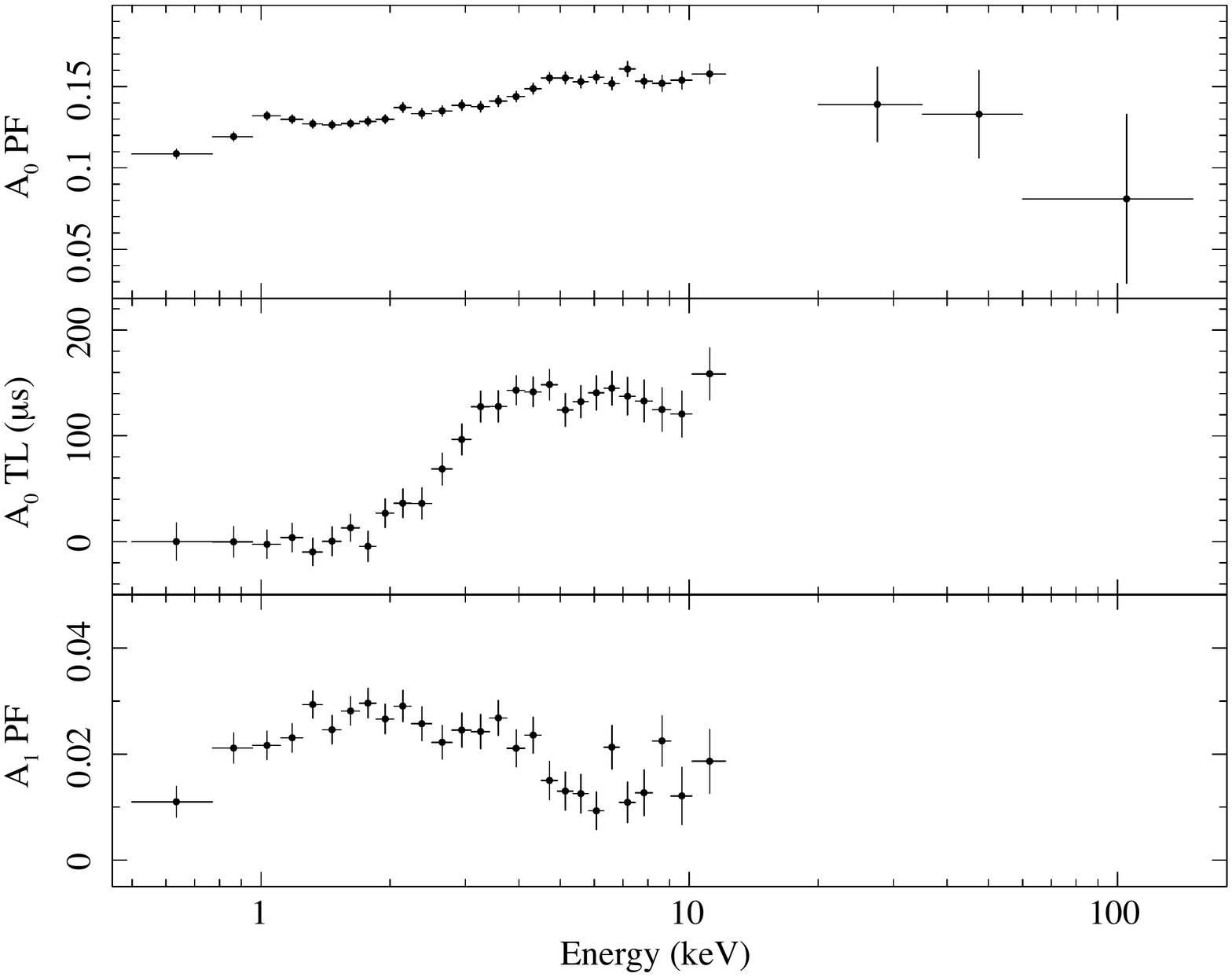, trim=1cm 1cm 1cm 0.7cm, scale=0.38}
\caption{Energy-dependent pulsed fraction (PF) for the fundamental, $A_0$, and first, $A_1$, harmonic and phase/time lag (TL) of the pulsed signal for the fundamental, $A_0$, harmonic \citep[see][for more details]{ferrigno14}. The data points between 0.5  and 11 keV are obtained combining the two \xmm\ observations, and the data between 20 and 150 keV are from IBIS/ISGRI. \igr\ shows a hard lag, normally not observed in other AMXPs (except for IGR J00291+5934).}
\label{fig:pf}
\end{figure}

In Fig. \ref{fig:pf}, we report the pulsed fraction of the fundamental and first harmonics in the 0.5 -- 11 keV and the phase/time lag of the fundamental harmonic obtained combining the two \xmm\ observations. The phase/time lag of the first harmonic is poorly determined, constant to zero, and not reported in this plot. The relative phase/time lags are expressed in microseconds as a function of energy compared to the averaged pulse profile \citep[see Eq. (1) in][and for more details]{ferrigno14}. The zero is arbitrarily taken at the lowest energy band. For ISGRI we derived the time averaged pulsed fraction (across the outburst) in three energy bands of the fundamental component considering the signal-to-noise of the pulse signals above 20 keV. These are derived from fitting a sinusoid to the ISGRI pulse profiles to determine the pulsed counts, converting the pulsed excess counts to photon fluxes and finally dividing this number by the total source flux as derived through ISGRI imaging. The ISGRI pulsed fraction value in the 20 -- 60 keV is $\sim14\%$, and connects well with the $\sim15\%$ near 10 keV in the \xmm\ data. 

For AMXPs, it was found that the low-energy pulses lag behind the high-energy pulses (soft phase/time lags), monotonically increasing with energy, and saturating at about 10 -- 20 keV \citep[see e.g.][]{cui98,ford00,gp05,Falanga05,i11}. This soft time lag has been interpreted as the result of photon delay due to down-scattering of hard X-ray photons in the relatively cold plasma of the disk or NS surface \citep{cui98,t02,falanga07}. On the other hand, \citet{pg03,gp05,ip09} suggested that the lags may be produced by a combination of different angular distributions of the radiation components, and differences in the emissivity pattern. On the contrary, \igr\ shows a hard lag, that is, low-energy pulsed photons arrive before the hard-energy pulsed photons. A similar trend in pulsed fraction and time lag has been observed for IGR J00291+5934, but only starting at higher energies from $\sim6$ keV \citep{Falanga05,Sanna17}. For \igr\ the thermal seed soft photons, coming from a larger emitting area, may up-scatter off hot electrons in the accretion column and arrive before the hard-energy photons. However, such a Compton-up scattering model is unlikely, since the lags are measured in the pulsed emission and the typical light-crossing time of the emission region is orders of magnitude smaller than the observed lags. It is more probable that the lags reflect variations in the emission pattern as a function of energy. Small deviations of the radiation angular distribution from the Lambert law induce rather large deviations in the pulse profile, leading therefore to lags \citep{pb06,lamb09}. Since the time lags are constant between 0.7 and 2 keV, this might suggest that these photons are coming from the disk. However, an increasing pulsed fraction in the range between $11$ and $14\%$ confirms rather the absence of the accretion disk at this energy range. We note that the maximum observed time delay, $\sim150 \ \mu$s, is comparable in absolute value with other AMXPs, indicating that they share most likely the same geometrical emission size or that the emission pattern has nearly the same energy gradient. The pulsed fraction of the first harmonic is between $1$ and $3\%$, while the time lags are consistent with zero due to large errors. The presence of the first harmonic may indicate that some pulsed emission is coming from the anti-polar cap, not being occulted by a disk, or that the emission pattern is not blackbody-like. 

\section{Properties of the type-I X-ray bursts} 
\label{sec:burst} 
During the 2013 outburst of \igr\ a total of three type-I X-ray bursts were detected (see Table \ref{tab:burst_tot}). The type-I X-ray bursts were separated by similar time intervals of $\Delta t_{\rm rec,1-2}=4.8$ d and $\Delta t_{\rm rec,2-3}=5.2$ d (see Fig. \ref{fig:lcr} and  Table \ref{tab:burst_tot}) and went off during the flat part of the outburst when the source was at the highest X-ray luminosity (i.e. away from the initial rise and final decay phases). 
\begin{table}[h] 
\caption{Observed type-I X-ray bursts during the outburst of \igr\ in 2013.}
\centering
\begin{tabular}{cccc} 
\hline 
Start  Time & Instrument & $\Delta t_{\rm rec}$ &Reference\tablefootmark{a} \\
(UTC) & & (d) &\\
\hline 
April 3 at 03:10:02 & MAXI & -- &[1]\\
April 7 at 22:18:05 & \swift & 4.8 &[2]\\
April 13 at 04:15:27 & \Integ & 5.2 & this paper\\
\hline  
\end{tabular}  
\tablefoot{ \tablefoottext{a}{[1] \citet{Serino13}; [2] \citet{papitto13a,Linares13}.}}
\label{tab:burst_tot} 
\end{table}  

\begin{figure}[h] 
\centering
\epsfig{figure=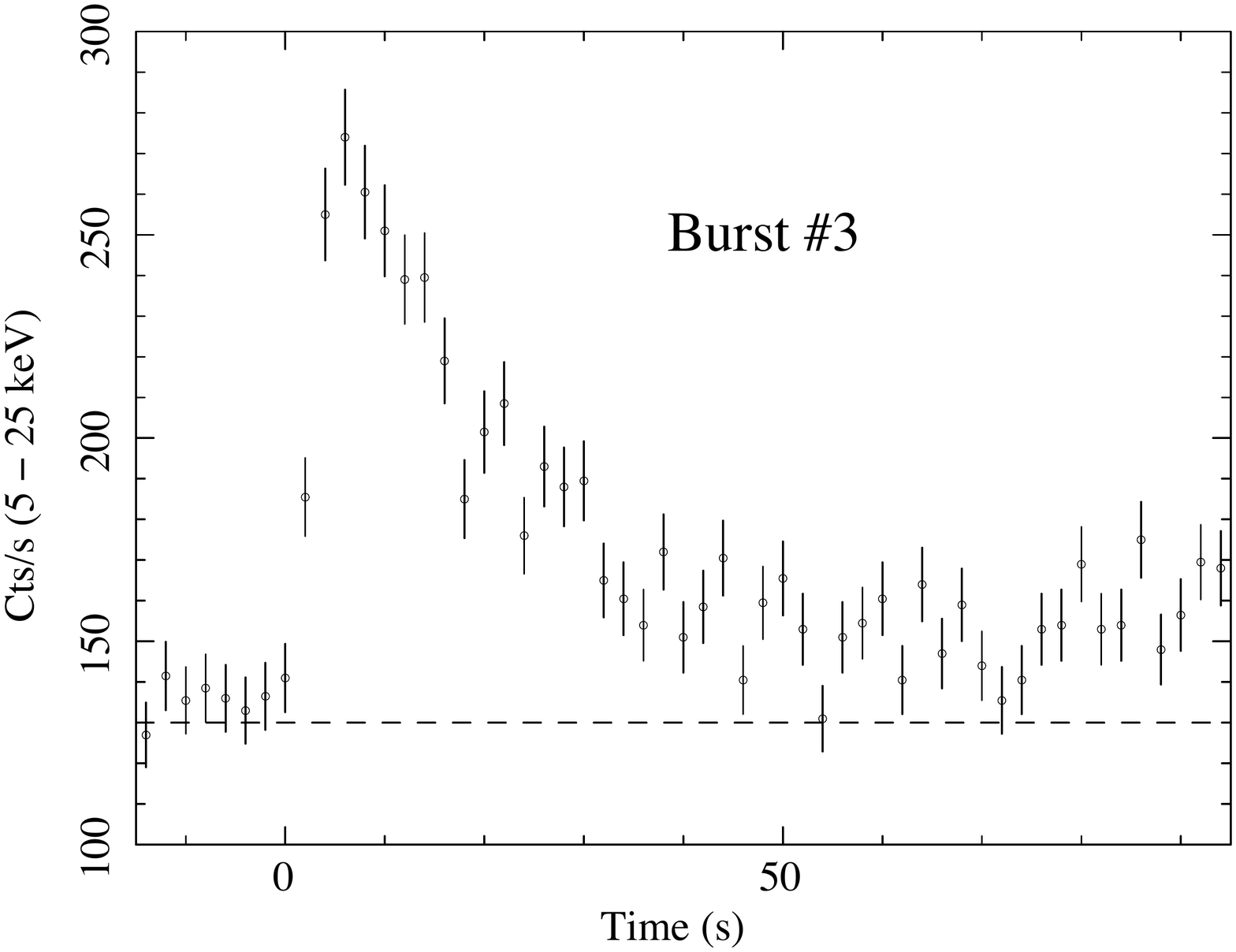, trim=1cm 0cm 3cm 2cm, scale=0.35}
\caption{The light curve of the type-I X-ray burst detected by JEM-X and reported for the first time in this paper. The burst start time was 56389.9292192~MJD. The JEM-X light curve was extracted in the 5 -- 25 keV energy range with a time bin of 2~s.}
\label{fig:burst}
\end{figure}

The source emission during the MAXI burst could not be analysed in detail because the data were contaminated by the nearby source GS~1826--238. The study of the second burst was reported by \citet{linares14,papitto13c}. Here we discuss in more detail the third burst that was not yet reported in literature. We discovered this event during a careful analysis of the \Integ\ data in revolution 1282. The ScWs 4, 5, and 6 of this revolution were affected by a high radiation background when the \Integ\ satellite was coming out from the Earth radiation belts and thus we had to specifically force the OSA software to skip the standard GTI selection to obtain the source light curve for this period. The burst was discovered in the ScW 4 and a zoom into the relevant part of the light curve is shown in Fig.~\ref{fig:burst}. 
The burst lasted $\sim90$ s and it showed a rise time of $\sim5$~s. The burst decay profile could be well fitted with an exponential function and the correspondingly derived e-folding time is $\tau_{\rm fit}=20\pm3$ s. The parameters of the JEM-X burst are thus well compatible with those measured by \citet{linares14} for the XRT burst (see their Table~2). As the spectral analysis of the JEM-X burst could not be carried out due to contamination from the radiation belts, we determined the burst persistent flux after 0.08 d from the JEM-X burst onset. The flux value is $F_{\rm pers}\approx3.6\times10^{-9}$ erg cm$^{-2}$ s$^{-1}$, corresponding to a luminosity of $L_{\rm pers,bol}\approx1.3\times 10^{37}$ erg s${}^{-1}$ (i.e. $3.4\%\ L_{\rm Edd}$, where $L_{\rm Edd}$ is the Eddington luminosity) close to the value reported by \citet{linares14}. However, since the persistent flux of the source is variable, we discuss later the results considering the flux range $F_{\rm pers}=(3-9)\times10^{-9}$ erg cm$^{-2}$ s$^{-1}$ for 0.11 d after the burst onset. If we assume that the total burst energy released during the JEM-X burst is comparable to that measured during the XRT burst, we obtain a total burst fluence of $f_{\rm b}=1.1\times10^{-6}$~erg~cm$^{-2}$ \citep{linares14}. As the light curve of the JEM-X burst does not show any clear evidence of a plateau at its peak, we conclude that most likely no photospheric radius expansion took place. We summarise in Table~\ref{tab:burst} all measured and extrapolated JEM-X burst parameters. 
\begin{table}[h] 
\caption{Parameters of the type-I X-ray burst observed by \Integ/JEM-X during the outburst of \igr\ in 2013.}
\centering
\begin{tabular}{ll} 
\hline 
\hline 
$\Delta t_{\rm burst}$ (s)  & $90\pm1$\\
$\Delta t_{\rm rise}$ (s) & $5\pm1$\\
$\tau_{\rm fit}$ (s) & $20\pm3$\\
$f_{\rm b}^*$ ($10^{-6}$ erg cm$^{-2}$) & $1.1\pm0.1$\\
$F_{\rm pers}^*$ ($10^{-9}$ erg cm$^{-2}$ s$^{-1}$) & $3.6\pm0.7$\\
\hline  
\end{tabular}  
\tablefoot{$^*$Extrapolated values.}
\label{tab:burst} 
\end{table} 

To constrain the nature of the thermonuclear burning that gave rise to the JEM-X burst, we first determine 
the local accretion rate per unit area onto the compact object at the time of the event as $\dot{m} = L_{\rm pers} (1+z) (4\pi R^2(GM/R_{\rm NS}))^{-1}$, that is, $\dot m \approx 7.3\times10^3$ g cm$^{-2}$ s$^{-1}$ (where the gravitational redshift is $1+ z = 1.31$ for a canonical NS with a mass $M_{\rm NS}=1.4M_\odot$ and a radius of $R_{\rm NS}=10$ km). We can then estimate the ignition depth at the onset of the burst with the equation $y_{\rm ign} = E_{\rm burst} (1+z)(4\pi R^2Q_{\rm nuc})^{-1}$, where the nuclear energy generated for pure helium (assuming a mean hydrogen mass fraction at ignition $\langle X\rangle=0$) is $Q_{\rm nuc}\approx 1.6+4\langle X\rangle\ {\rm MeV/nucleon}\approx1.6\ {\rm MeV/nucleon}$ and for solar abundances (assuming $\langle X\rangle=0.7$) is $Q_{\rm nuc}\approx4.4\ {\rm MeV/nucleon}$ \citep[][and references therein]{galloway04}. We note that the considered values $Q_{\rm nuc}$ include losses owing to neutrino emission as detailed in \citet{fujimoto87}. Once the ignition depth is known, the recurrence time between the bursts can be calculated by using the equation $\Delta t_{\rm rec} =(y_{\rm ign} /\dot{m})(1 + z)$. We obtain $y_{\rm ign} =2.7\times10^8$ g cm${}^{-2}$ and $\Delta t_{\rm rec}\sim 5.6$~d in the case of helium burning, while $y_{\rm ign} = 10^8$ g cm${}^{-2}$ and $\Delta t_{\rm rec}\sim 2.1$~d in the 
hydrogen burning case. 

These results indicate that the JEM-X event could be more likely associated to a helium burst. However, to properly disentangle between the cases of helium and hydrogen bursts, we compare the total accreted matter between two subsequent bursts with the amount of fuel liberated during these events. From Fig.~\ref{fig:lcr} we estimate that the mean persistent fluxes (0.4 -- 250 keV energy band) between the observed bursts are $F_{\rm pers,1-2}=(1.6\pm0.1)\times10^{-9}$ erg cm${}^{-2}$ s${}^{-1}$ and $F_{\rm pers,2-3}=(2.0\pm0.1)\times10^{-9}$ erg cm${}^{-2}$ s${}^{-1}$, respectively. These correspond to a persistent luminosity of $L_{\rm X}\approx(6-7)\times 10^{36}$ erg s${}^{-1}$ (or $(1.5-2)\%\ L_{\rm Edd}$). The mass accretion rate is thus $\dot{M}=(6-8)\times10^{16}$ g s$^{-1}$, as calculated through the standard accretion equation $L_{\rm X}=GM_{\rm NS}\dot{M}/R_{\rm NS}$ \citep[see, e.g.][]{frank02}. The estimated total mass accreted between the type-I X-ray bursts is thus $\Delta M = \dot{M} \Delta t \approx (2.6-3.6)\times 10^{21}$ g, using $\Delta t=4.8$~d and $5.2$~d. The amount of fuel liberated during a thermonuclear burst can be estimated as $\Delta M_{\rm He}=E_{\rm burst}/\varepsilon_{\rm He}\approx 2.4 \times 10^{21}$~g and $\Delta M_{\rm H}=E_{\rm burst}/\varepsilon_{\rm H}\approx 4.9 \times 10^{20}$~g. Here $\varepsilon_{\rm He} \approx1.6 \times 10^{18}$ erg g$^{-1}$ is the total available nuclear energy for the transformation of pure helium into iron-peak elements, and $\varepsilon_{\rm H} \approx8 \times 10^{18}$ erg g$^{-1}$ is the total available nuclear energy for the transformation of pure hydrogen into iron-peak elements. As can be immediately noticed, $\Delta M$ and $\Delta M_{\rm He}$ are the closest values. Taking all the results together, we conclude that the JEM-X burst was triggered by unstable helium burning, after all accreted hydrogen was exhausted by the steady burning prior to the burst. If we consider the lower and upper burst persistent flux level of $F_{\rm pers}=(3-9)\times10^{-9}$ erg cm$^{-2}$ s$^{-1}$, this corresponds to a mixed helium/hydrogen burst ignition regime triggered by thermally unstable helium ignition \citep[see][]{Strohmayer03,linares14}, with a recurrence time, for solar composition, of between $\Delta t_{\rm rec}=(1$ and $2.5)$ d and for helium between $\Delta t_{\rm rec}=(2.3$ and $6.7)$ d. Such a short recurrence time for solar composition has not been observed during our data set including also the 131 ks \Integ\ continuous data. For helium composition instead, the range is more suitable with the observed value. We note that our conclusion  is also in agreement with the light curve profiles of both XRT and JEM-X bursts, as helium bursts typically show a fast rise time of a few seconds at the most \citep{lewin93,galloway08}. The probabilities of detecting a burst with $\Delta t_{\rm rec}=5.6$ d in the intervals (56377.5 -- 56385.2) MJD, (56385.2 -- 56390) MJD, and (56390 -- 56397) MJD are respectively $P_1=54\%$, $P_2=88\%$, and $P_3=67\%$. Instead for the lower limit recurrence time, $\Delta t_{\rm rec}=2.3$ d, the probabilities decrease as $P_1=39\%$, $P_2=83\%$, and $P_3=56\%$. Therefore, there is higher probability of missing another type-I X-ray burst at the beginning and at the end of the outburst. 

\section{Summary and discussion} 
\label{sec:summary} 
In this paper, we reported for the first time on a detailed analysis of the \Integ\ data collected during the 2013 outburst of the transitional AMXP \igr. The source is known for having displayed a peculiarly prominent timing and spectral variability in the X-ray domain during a full accretion episode, on timescales as short as fractions of a second. This variability is not usually found in classical AMXPs, which have never before shown (to the best of our knowledge) evidence of transitions between phases of rotationally and accretion-powered pulsations. PSR J1023+0038 and XSS J1227.0--4859 are another two transient AMXPs showing a fast flux variability, but they have never been observed in full outburst \citep{linares14b}. In addition, their variability is accompanied by flaring and dip states, that \igr\ does not show \citep{demartino10,Shahbaz15}.

At odds with the peculiar timing and spectral variability displayed by \igr\ on short timescales, the properties of the source X-ray emission on timescales of a few days seem to be remarkably similar to those of other AMXPs in outburst. The overall outburst profile is also closely reminiscent of that observed from AMXPs, featuring a faster rise ($\sim$ 1 -- 2~days) and a slower decay (several days) \citep[see e.g. ][]{Gilfanov98,gp05,Powell07,falanga12}. The other two transient AMXPs show highly variable flat X-ray light curves with peculiar low luminosities of $\sim10^{33}$ erg s$^{-1}$ and flares reaching $\sim10^{34}$ erg s$^{-1}$ \citep{demartino10,bogdanov15}. 

The \Integ\ data permitted us to carry out an analysis of the averaged broad-band spectral properties of the source, covering the 0.4 -- 250~keV energy range. We showed in Sect.~\ref{sec:spe} that a Comptonisation model can describe the energy distribution of the X-ray photons from \igr. At the contrary of other AMXPs this source displays a harder spectrum, due to larger emitting area and smaller seed photon temperature \citep[see e.g. ][]{gdb02,gp05,falanga12}. The X-ray spectra of the other two transient AMXPs are described by an absorbed power law and are softer, showing spectral slopes in the range $\Gamma=(1.6-1.7)$ \citep{demartino10,bogdanov15}.

We improved the \xmm\ ephemeris reported by \citet{papitto13c}, finding a slightly lower spin frequency. Thanks to this result, we have now a consistent alignment in the folding procedure of the two 2 -- 10 keV \xmm\ pulse-phase distributions and, for the first time, the detection of the millisecond pulsations using the TTE data with \Fermi/GBM. For the 20 -- 60 keV \Fermi/GBM band we detected pulsed emission at a $5.9\sigma$ confidential level with a pulse shape fully consistent with that observed by ISGRI 20 -- 60 keV. The pulse profiles of \igr\ share many similarities with those of other AMXPs, as their shape is nearly sinusoidal at all energies \citep{patruno12}. The hard time lags of the pulsed emission likely indicate that the emission pattern from the hotspot has a peculiar energy dependence different from that of other AMXPs \citep{Falanga05}. Alternatively, the contribution of the secondary cap changes the pulse profile to affect the sign of the lags. It would be interesting to compare the hard lags of \igr\ with those of the other two transient AMXPs, since they have spin frequencies double that of \igr\ \citep[][]{Archibald13,demartino14}.

We reported on the discovery of a previously undetected thermonuclear burst from the source, caught by the JEM-X monitors at the beginning of the revolution 1282 when the \Integ\ satellite was coming out from the Earth radiation belts. Even though we could not perform a detailed spectral analysis of the event, the similarity with a previous burst detected by \swift/XRT allowed us to demonstrate that type-I X-ray bursts from \igr\ are most likely triggered by unstable helium burning after the exhaustion of all accreted hydrogen on the NS surface. This conclusion is compatible with both the characteristics of the burst profiles recorded from the source and with their measured recurrence time of roughly 5~days. \igr\ is the only transient AMXP exhibiting a type-I X-ray burst so far. The donor stars hosted in the other two transient AMXPs were identified to be G-type stars with a mass ranging between (0.2 -- 0.4) M$_\odot$ \citep{Archibald13,demartino14}. This is in agreement with evolutionary expectations, which predict that AMXPs with an orbital period in the hour range (for these two transient AMXPs this is $\sim(5-7)$ hr or half of \igr\ period \citep{Archibald13,demartino14}) should host a hydrogen-rich companion star \citep{deloye03}. These systems are also expected exhibit pure helium type-I X-ray bursts \citep[see e.g.][] {galloway06,watts06,falanga07,galloway07,ferrigno11,defalco17}. We estimated that the burst recurrence time, for a persistent luminosity of $L=10^{34}$ erg s$^{-1}$ and assuming that the helium burst is similar to that exhibited by \igr, is $\Delta_{\rm rec}=20$ yr. The transition between radio and X-ray phases may further delay the occurrence of the type-I X-ray bursts.

\begin{acknowledgements} 
This research was financed by the Swiss National Science Foundation project 200021\_149865. VDF and MF  acknowledge the Department Physics, University of Basel, specially Friedrich-K. Thielemann and the International Space Science Institute in Bern for their support. VdF is grateful to the ISDC Data Center for Astrophysics in Versoix for their hospitality, where part of this work has been carried out. JP thanks the Academy of Finland (grant 268740), the Foundations' Professor Pool and the Finnish Cultural Foundation for support.  
\end{acknowledgements} 

\bibliographystyle{aa}
\bibliography{references}

\begin{thebibliography}{94}
\expandafter\ifx\csname natexlab\endcsname\relax\def\natexlab#1{#1}\fi

\bibitem[{{Alpar} {et~al.}(1982){Alpar}, {Cheng}, {Ruderman}, \&
  {Shaham}}]{Alpar82}
{Alpar}, M.~A., {Cheng}, A.~F., {Ruderman}, M.~A., \& {Shaham}, J. 1982, \nat,
  300, 728

\bibitem[{{Archibald} {et~al.}(2013){Archibald}, {Kaspi}, {Hessels},
  {Stappers}, {Janssen}, \& {Lyne}}]{Archibald13}
{Archibald}, A.~M., {Kaspi}, V.~M., {Hessels}, J.~W.~T., {et~al.} 2013, ArXiv
  e-prints

\bibitem[{{Archibald} {et~al.}(2009){Archibald}, {Stairs}, {Ransom}, {Kaspi},
  {Kondratiev}, {Lorimer}, {McLaughlin}, {Boyles}, {Hessels}, {Lynch}, {van
  Leeuwen}, {Roberts}, {Jenet}, {Champion}, {Rosen}, {Barlow}, {Dunlap}, \&
  {Remillard}}]{archibald09}
{Archibald}, A.~M., {Stairs}, I.~H., {Ransom}, S.~M., {et~al.} 2009, Science,
  324, 1411

\bibitem[{{Arnaud}(1996)}]{arnaud96}
{Arnaud}, K.~A. 1996, in Astronomical Society of the Pacific Conference Series,
  Vol. 101, Astronomical Data Analysis Software and Systems V, ed. G.~H.
  {Jacoby} \& J.~{Barnes}, 17

\bibitem[{{Atwood} {et~al.}(2009){Atwood}, {Abdo}, {Ackermann}, {Althouse},
  {Anderson}, {Axelsson}, {Baldini}, {Ballet}, {Band}, {Barbiellini}, \&
  et~al.}]{atwood09}
{Atwood}, W.~B., {Abdo}, A.~A., {Ackermann}, M., {et~al.} 2009, \apj, 697, 1071

\bibitem[{{Bassa} {et~al.}(2014){Bassa}, {Patruno}, {Hessels}, {Keane},
  {Monard}, {Mahony}, {Bogdanov}, {Corbel}, {Edwards}, {Archibald}, {Janssen},
  {Stappers}, \& {Tendulkar}}]{bassa14}
{Bassa}, C.~G., {Patruno}, A., {Hessels}, J.~W.~T., {et~al.} 2014, \mnras, 441,
  1825

\bibitem[{{Becker} {et~al.}(2003){Becker}, {Swartz}, {Pavlov}, {Elsner},
  {Grindlay}, {Mignani}, {Tennant}, {Backer}, {Pulone}, {Testa}, \&
  {Weisskopf}}]{Becker03}
{Becker}, W., {Swartz}, D.~A., {Pavlov}, G.~G., {et~al.} 2003, \apj, 594, 798

\bibitem[{{Bisnovatyi-Kogan} \& {Komberg}(1974)}]{bk74}
{Bisnovatyi-Kogan}, G.~S. \& {Komberg}, B.~V. 1974, \sovast, 18, 217

\bibitem[{{Bissaldi} {et~al.}(2009){Bissaldi}, {von Kienlin}, {Lichti},
  {Steinle}, {Bhat}, {Briggs}, {Fishman}, {Hoover}, {Kippen}, {Krumrey},
  {Gerlach}, {Connaughton}, {Diehl}, {Greiner}, {van der Horst}, {Kouveliotou},
  {McBreen}, {Meegan}, {Paciesas}, {Preece}, \& {Wilson-Hodge}}]{bissaldi09}
{Bissaldi}, E., {von Kienlin}, A., {Lichti}, G., {et~al.} 2009, Experimental
  Astronomy, 24, 47

\bibitem[{{Bogdanov} {et~al.}(2014){Bogdanov}, {Archibald}, {Bassa}, {Deller},
  {Halpern}, {Heald}, {Hessels}, {Janssen}, {Lyne}, {Moldon}, {Paragi},
  {Patruno}, {Perera}, {Stappers}, {Tendulkar}, {D'Angelo}, \&
  {Wijnands}}]{bogdanov14}
{Bogdanov}, S., {Archibald}, A.~M., {Bassa}, C., {et~al.} 2014, ArXiv e-prints

\bibitem[{{Bogdanov} {et~al.}(2015){Bogdanov}, {Archibald}, {Bassa}, {Deller},
  {Halpern}, {Heald}, {Hessels}, {Janssen}, {Lyne}, {Mold{\'o}n}, {Paragi},
  {Patruno}, {Perera}, {Stappers}, {Tendulkar}, {D'Angelo}, \&
  {Wijnands}}]{bogdanov15}
{Bogdanov}, S., {Archibald}, A.~M., {Bassa}, C., {et~al.} 2015, \apj, 806, 148

\bibitem[{{Bozzo} {et~al.}(2016){Bozzo}, {Pjanka}, {Romano}, {Papitto},
  {Ferrigno}, {Motta}, {Zdziarski}, {Pintore}, {Di Salvo}, {Burderi},
  {Lazzati}, {Ponti}, \& {Pavan}}]{bozzo16}
{Bozzo}, E., {Pjanka}, P., {Romano}, P., {et~al.} 2016, \aap, 589, A42

\bibitem[{{Buccheri} {et~al.}(1983){Buccheri}, {Bennett}, {Bignami}, {Bloemen},
  {Boriakoff}, {Caraveo}, {Hermsen}, {Kanbach}, {Manchester}, {Masnou},
  {Mayer-Hasselwander}, {Ozel}, {Paul}, {Sacco}, {Scarsi}, \&
  {Strong}}]{buccheri1983}
{Buccheri}, R., {Bennett}, K., {Bignami}, G.~F., {et~al.} 1983, \aap, 128, 245

\bibitem[{{Burrows} {et~al.}(2005){Burrows}, {Hill}, {Nousek}, {Kennea},
  {Wells}, {Osborne}, {Abbey}, {Beardmore}, {Mukerjee}, {Short}, {Chincarini},
  {Campana}, {Citterio}, {Moretti}, {Pagani}, {Tagliaferri}, {Giommi},
  {Capalbi}, {Tamburelli}, {Angelini}, {Cusumano}, {Br{\"a}uninger}, {Burkert},
  \& {Hartner}}]{burrows05}
{Burrows}, D.~N., {Hill}, J.~E., {Nousek}, J.~A., {et~al.} 2005, \ssr, 120, 165

\bibitem[{{Cohn} {et~al.}(2013){Cohn}, {Lugger}, {Bogdanov}, {Heinke}, {Van Den
  Berg}, \& {Sivakoff}}]{Cohn13}
{Cohn}, H.~N., {Lugger}, P.~M., {Bogdanov}, S., {et~al.} 2013, The Astronomer's
  Telegram, 5031

\bibitem[{{Courvoisier} {et~al.}(2003){Courvoisier}, {Walter}, {Beckmann},
  {Dean}, {Dubath}, {Hudec}, {Kretschmar}, {Mereghetti}, {Montmerle},
  {Mowlavi}, {Paltani}, {Preite Martinez}, {Produit}, {Staubert}, {Strong},
  {Swings}, {Westergaard}, {White}, {Winkler}, \& {Zdziarski}}]{c03}
{Courvoisier}, T.~J.-L., {Walter}, R., {Beckmann}, V., {et~al.} 2003, \aap,
  411, L53

\bibitem[{{Cui} {et~al.}(1998){Cui}, {Morgan}, \& {Titarchuk}}]{cui98}
{Cui}, W., {Morgan}, E.~H., \& {Titarchuk}, L.~G. 1998, \apjl, 504, L27

\bibitem[{{De Falco} {et~al.}(2017){De Falco}, {Kuiper}, {Bozzo}, {Galloway},
  {Poutanen}, {Ferrigno}, {Stella}, \& {Falanga}}]{defalco17}
{De Falco}, V., {Kuiper}, L., {Bozzo}, E., {et~al.} 2017, \aap, 599, A88

\bibitem[{{de Martino} {et~al.}(2013){de Martino}, {Belloni}, {Falanga},
  {Papitto}, {Motta}, {Pellizzoni}, {Evangelista}, {Piano}, {Masetti},
  {Bonnet-Bidaud}, {Mouchet}, {Mukai}, \& {Possenti}}]{DeMartino13}
{de Martino}, D., {Belloni}, T., {Falanga}, M., {et~al.} 2013, \aap, 550, A89

\bibitem[{{de Martino} {et~al.}(2014){de Martino}, {Casares}, {Mason},
  {Buckley}, {Kotze}, {Bonnet-Bidaud}, {Mouchet}, {Coppejans}, \&
  {Gulbis}}]{demartino14}
{de Martino}, D., {Casares}, J., {Mason}, E., {et~al.} 2014, \mnras, 444, 3004

\bibitem[{{de Martino} {et~al.}(2010){de Martino}, {Falanga}, {Bonnet-Bidaud},
  {Belloni}, {Mouchet}, {Masetti}, {Andruchow}, {Cellone}, {Mukai}, \&
  {Matt}}]{demartino10}
{de Martino}, D., {Falanga}, M., {Bonnet-Bidaud}, J.-M., {et~al.} 2010, \aap,
  515, A25

\bibitem[{{Deloye} \& {Bildsten}(2003)}]{deloye03}
{Deloye}, C.~J. \& {Bildsten}, L. 2003, \apj, 598, 1217

\bibitem[{{Di Salvo} \& {Burderi}(2003)}]{DiSalvo03}
{Di Salvo}, T. \& {Burderi}, L. 2003, \aap, 397, 723

\bibitem[{{Dickey} \& {Lockman}(1990)}]{dickey90}
{Dickey}, J.~M. \& {Lockman}, F.~J. 1990, \araa, 28, 215

\bibitem[{{Eckert} {et~al.}(2013){Eckert}, {Del Santo}, {Bazzano}, {Watanabe},
  {Paizis}, {Bozzo}, {Ferrigno}, {Caballero}, {Sidoli}, \& {Kuiper}}]{eckert13}
{Eckert}, D., {Del Santo}, M., {Bazzano}, A., {et~al.} 2013, The Astronomer's
  Telegram, 4925

\bibitem[{{Evans} {et~al.}(2009){Evans}, {Beardmore}, {Page}, {Osborne},
  {O'Brien}, {Willingale}, {Starling}, {Burrows}, {Godet}, {Vetere}, {Racusin},
  {Goad}, {Wiersema}, {Angelini}, {Capalbi}, {Chincarini}, {Gehrels}, {Kennea},
  {Margutti}, {Morris}, {Mountford}, {Pagani}, {Perri}, {Romano}, \&
  {Tanvir}}]{evans09}
{Evans}, P.~A., {Beardmore}, A.~P., {Page}, K.~L., {et~al.} 2009, \mnras, 397,
  1177

\bibitem[{{Falanga} {et~al.}(2005{\natexlab{a}}){Falanga}, {Bonnet-Bidaud},
  {Poutanen}, {Farinelli}, {Martocchia}, {Goldoni}, {Qu}, {Kuiper}, \&
  {Goldwurm}}]{mfb05}
{Falanga}, M., {Bonnet-Bidaud}, J.~M., {Poutanen}, J., {et~al.}
  2005{\natexlab{a}}, \aap, 436, 647

\bibitem[{{Falanga} {et~al.}(2005{\natexlab{b}}){Falanga}, {Kuiper},
  {Poutanen}, {Bonning}, {Hermsen}, {di Salvo}, {Goldoni}, {Goldwurm}, {Shaw},
  \& {Stella}}]{Falanga05}
{Falanga}, M., {Kuiper}, L., {Poutanen}, J., {et~al.} 2005{\natexlab{b}}, \aap,
  444, 15

\bibitem[{{Falanga} {et~al.}(2011){Falanga}, {Kuiper}, {Poutanen}, {Galloway},
  {Bonning}, {Bozzo}, {Goldwurm}, {Hermsen}, \& {Stella}}]{falanga11}
{Falanga}, M., {Kuiper}, L., {Poutanen}, J., {et~al.} 2011, \aap, 529, A68

\bibitem[{{Falanga} {et~al.}(2012){Falanga}, {Kuiper}, {Poutanen}, {Galloway},
  {Bozzo}, {Goldwurm}, {Hermsen}, \& {Stella}}]{falanga12}
{Falanga}, M., {Kuiper}, L., {Poutanen}, J., {et~al.} 2012, \aap, 545, A26

\bibitem[{{Falanga} {et~al.}(2007){Falanga}, {Poutanen}, {Bonning}, {Kuiper},
  {Bonnet-Bidaud}, {Goldwurm}, {Hermsen}, \& {Stella}}]{mfc07}
{Falanga}, M., {Poutanen}, J., {Bonning}, E.~W., {et~al.} 2007, \aap, 464, 1069

\bibitem[{{Falanga} \& {Titarchuk}(2007)}]{falanga07}
{Falanga}, M. \& {Titarchuk}, L. 2007, \apj, 661, 1084

\bibitem[{{Ferrigno} {et~al.}(2011){Ferrigno}, {Bozzo}, {Falanga}, {Stella},
  {Campana}, {Belloni}, {Israel}, {Pavan}, {Kuulkers}, \&
  {Papitto}}]{ferrigno11}
{Ferrigno}, C., {Bozzo}, E., {Falanga}, M., {et~al.} 2011, \aap, 525, A48

\bibitem[{{Ferrigno} {et~al.}(2014){Ferrigno}, {Bozzo}, {Papitto}, {Rea},
  {Pavan}, {Campana}, {Wieringa}, {Filipovi{\'c}}, {Falanga}, \&
  {Stella}}]{ferrigno14}
{Ferrigno}, C., {Bozzo}, E., {Papitto}, A., {et~al.} 2014, \aap, 567, A77

\bibitem[{{Ford}(2000)}]{ford00}
{Ford}, E.~C. 2000, \apjl, 535, L119

\bibitem[{{Frank} {et~al.}(2002){Frank}, {King}, \& {Raine}}]{frank02}
{Frank}, J., {King}, A., \& {Raine}, D.~J. 2002, {Accretion Power in
  Astrophysics: Third Edition}, 398

\bibitem[{{Fujimoto} {et~al.}(1987){Fujimoto}, {Sztajno}, {Lewin}, \& {van
  Paradijs}}]{fujimoto87}
{Fujimoto}, M.~Y., {Sztajno}, M., {Lewin}, W.~H.~G., \& {van Paradijs}, J.
  1987, \apj, 319, 902

\bibitem[{{Galloway}(2008)}]{galloway08}
{Galloway}, D. 2008, in American Institute of Physics Conference Series, Vol.
  983, 40 Years of Pulsars: Millisecond Pulsars, Magnetars and More, ed.
  C.~{Bassa}, Z.~{Wang}, A.~{Cumming}, \& V.~M. {Kaspi}, 510--518

\bibitem[{{Galloway} \& {Cumming}(2006)}]{galloway06}
{Galloway}, D.~K. \& {Cumming}, A. 2006, \apj, 652, 559

\bibitem[{{Galloway} {et~al.}(2004){Galloway}, {Cumming}, {Kuulkers},
  {Bildsten}, {Chakrabarty}, \& {Rothschild}}]{galloway04}
{Galloway}, D.~K., {Cumming}, A., {Kuulkers}, E., {et~al.} 2004, \apj, 601, 466

\bibitem[{{Galloway} {et~al.}(2007){Galloway}, {Morgan}, {Krauss}, {Kaaret}, \&
  {Chakrabarty}}]{galloway07}
{Galloway}, D.~K., {Morgan}, E.~H., {Krauss}, M.~I., {Kaaret}, P., \&
  {Chakrabarty}, D. 2007, \apjl, 654, L73

\bibitem[{{Gierli{\'n}ski} {et~al.}(2002){Gierli{\'n}ski}, {Done}, \&
  {Barret}}]{gdb02}
{Gierli{\'n}ski}, M., {Done}, C., \& {Barret}, D. 2002, \mnras, 331, 141

\bibitem[{{Gierli{\'n}ski} \& {Poutanen}(2005)}]{gp05}
{Gierli{\'n}ski}, M. \& {Poutanen}, J. 2005, \mnras, 359, 1261

\bibitem[{{Gilfanov} {et~al.}(1998){Gilfanov}, {Revnivtsev}, {Sunyaev}, \&
  {Churazov}}]{Gilfanov98}
{Gilfanov}, M., {Revnivtsev}, M., {Sunyaev}, R., \& {Churazov}, E. 1998, \aap,
  338, L83

\bibitem[{{Goldwurm} {et~al.}(2003){Goldwurm}, {David}, {Foschini}, {Gros},
  {Laurent}, {Sauvageon}, {Bird}, {Lerusse}, \& {Produit}}]{gold03}
{Goldwurm}, A., {David}, P., {Foschini}, L., {et~al.} 2003, \aap, 411, L223

\bibitem[{{Gros} {et~al.}(2003){Gros}, {Goldwurm}, {Cadolle-Bel}, {Goldoni},
  {Rodriguez}, {Foschini}, {Del Santo}, \& {Blay}}]{gros03}
{Gros}, A., {Goldwurm}, A., {Cadolle-Bel}, M., {et~al.} 2003, \aap, 411, L179

\bibitem[{{Harris}(1996)}]{Harris96}
{Harris}, W.~E. 1996, \aj, 112, 1487

\bibitem[{{Heinke} {et~al.}(2013){Heinke}, {Bahramian}, {Wijnands}, \&
  {Altamirano}}]{heinke13}
{Heinke}, C.~O., {Bahramian}, A., {Wijnands}, R., \& {Altamirano}, D. 2013, The
  Astronomer's Telegram, 4927

\bibitem[{{Homan} \& {Pooley}(2013)}]{homan13}
{Homan}, J. \& {Pooley}, D. 2013, The Astronomer's Telegram, 5045

\bibitem[{{Ibragimov} {et~al.}(2011){Ibragimov}, {Kajava}, \& {Poutanen}}]{i11}
{Ibragimov}, A., {Kajava}, J.~J.~E., \& {Poutanen}, J. 2011, \mnras, 415, 1864

\bibitem[{{Ibragimov} \& {Poutanen}(2009)}]{ip09}
{Ibragimov}, A. \& {Poutanen}, J. 2009, \mnras, 400, 492

\bibitem[{{Jansen} {et~al.}(2001){Jansen}, {Lumb}, {Altieri}, {Clavel}, {Ehle},
  {Erd}, {Gabriel}, {Guainazzi}, {Gondoin}, {Much}, {Munoz}, {Santos},
  {Schartel}, {Texier}, \& {Vacanti}}]{jansen01}
{Jansen}, F., {Lumb}, D., {Altieri}, B., {et~al.} 2001, \aap, 365, L1

\bibitem[{{Kaastra} \& {Bleeker}(2016)}]{Kaastra2016}
{Kaastra}, J.~S. \& {Bleeker}, J.~A.~M. 2016, \aap, 587, A151

\bibitem[{{Kalberla} {et~al.}(2005){Kalberla}, {Burton}, {Hartmann}, {Arnal},
  {Bajaja}, {Morras}, \& {P{\"o}ppel}}]{kalberla05}
{Kalberla}, P.~M.~W., {Burton}, W.~B., {Hartmann}, D., {et~al.} 2005, \aap,
  440, 775

\bibitem[{{Lamb} {et~al.}(2009){Lamb}, {Boutloukos}, {Van Wassenhove},
  {Chamberlain}, {Lo}, {Clare}, {Yu}, \& {Miller}}]{lamb09}
{Lamb}, F.~K., {Boutloukos}, S., {Van Wassenhove}, S., {et~al.} 2009, \apj,
  706, 417

\bibitem[{{Lebrun} {et~al.}(2003){Lebrun}, {Leray}, {Lavocat}, {Cr{\'e}tolle},
  {Arqu{\`e}s}, {Blondel}, {Bonnin}, {Bou{\`e}re}, {Cara}, {Chaleil}, {Daly},
  {Desages}, {Dzitko}, {Horeau}, {Laurent}, {Limousin}, {Mathy}, {Mauguen},
  {Meignier}, {Molini{\'e}}, {Poindron}, {Rouger}, {Sauvageon}, \&
  {Tourrette}}]{lebrun03}
{Lebrun}, F., {Leray}, J.~P., {Lavocat}, P., {et~al.} 2003, \aap, 411, L141

\bibitem[{{Lewin} {et~al.}(1993){Lewin}, {van Paradijs}, \& {Taam}}]{lewin93}
{Lewin}, W.~H.~G., {van Paradijs}, J., \& {Taam}, R.~E. 1993, \ssr, 62, 223

\bibitem[{{Linares}(2013)}]{Linares13}
{Linares}, M. 2013, The Astronomer's Telegram, 4960

\bibitem[{{Linares}(2014)}]{linares14b}
{Linares}, M. 2014, \apj, 795, 72

\bibitem[{{Linares} {et~al.}(2014){Linares}, {Bahramian}, {Heinke}, {Wijnands},
  {Patruno}, {Altamirano}, {Homan}, {Bogdanov}, \& {Pooley}}]{linares14}
{Linares}, M., {Bahramian}, A., {Heinke}, C., {et~al.} 2014, \mnras, 438, 251

\bibitem[{{Lund} {et~al.}(2003){Lund}, {Budtz-J{\o}rgensen}, {Westergaard},
  {Brandt}, {Rasmussen}, {Hornstrup}, {Oxborrow}, {Chenevez}, {Jensen},
  {Laursen}, {Andersen}, {Mogensen}, {Rasmussen}, {Om{\o}}, {Pedersen},
  {Polny}, {Andersson}, {Andersson}, {K{\"a}m{\"a}r{\"a}inen}, {Vilhu},
  {Huovelin}, {Maisala}, {Morawski}, {Juchnikowski}, {Costa}, {Feroci},
  {Rubini}, {Rapisarda}, {Morelli}, {Carassiti}, {Frontera}, {Pelliciari},
  {Loffredo}, {Mart{\'{\i}}nez N{\'u}{\~n}ez}, {Reglero}, {Velasco}, {Larsson},
  {Svensson}, {Zdziarski}, {Castro-Tirado}, {Attina}, {Goria}, {Giulianelli},
  {Cordero}, {Rezazad}, {Schmidt}, {Carli}, {Gomez}, {Jensen}, {Sarri},
  {Tiemon}, {Orr}, {Much}, {Kretschmar}, \& {Schnopper}}]{lund03}
{Lund}, N., {Budtz-J{\o}rgensen}, C., {Westergaard}, N.~J., {et~al.} 2003,
  \aap, 411, L231

\bibitem[{{Manchester} {et~al.}(2005){Manchester}, {Hobbs}, {Teoh}, \&
  {Hobbs}}]{Manchester05}
{Manchester}, R.~N., {Hobbs}, G.~B., {Teoh}, A., \& {Hobbs}, M. 2005, \aj, 129,
  1993

\bibitem[{{Meegan} {et~al.}(2009){Meegan}, {Lichti}, {Bhat}, {Bissaldi},
  {Briggs}, {Connaughton}, {Diehl}, {Fishman}, {Greiner}, {Hoover}, {van der
  Horst}, {von Kienlin}, {Kippen}, {Kouveliotou}, {McBreen}, {Paciesas},
  {Preece}, {Steinle}, {Wallace}, {Wilson}, \& {Wilson-Hodge}}]{meegan09}
{Meegan}, C., {Lichti}, G., {Bhat}, P.~N., {et~al.} 2009, \apj, 702, 791

\bibitem[{{Monard} \& {Kuulkers}(2013)}]{Monrad13}
{Monard}, L.~A.~G. \& {Kuulkers}, E. 2013, The Astronomer's Telegram, 4964

\bibitem[{{Pallanca} {et~al.}(2013{\natexlab{a}}){Pallanca}, {Dalessandro},
  {Ferraro}, {Lanzoni}, \& {Beccari}}]{Pallanca13b}
{Pallanca}, C., {Dalessandro}, E., {Ferraro}, F.~R., {Lanzoni}, B., \&
  {Beccari}, G. 2013{\natexlab{a}}, \apj, 773, 122

\bibitem[{{Pallanca} {et~al.}(2013{\natexlab{b}}){Pallanca}, {Dalessandro},
  {Ferraro}, {Lanzoni}, \& {Beccari}}]{Pallanca13a}
{Pallanca}, C., {Dalessandro}, E., {Ferraro}, R.~F., {Lanzoni}, B., \&
  {Beccari}, G. 2013{\natexlab{b}}, The Astronomer's Telegram, 5003

\bibitem[{{Papitto} {et~al.}(2013{\natexlab{a}}){Papitto}, {Bozzo}, {Ferrigno},
  {Pavan}, {Romano}, \& {Campana}}]{papitto13a}
{Papitto}, A., {Bozzo}, E., {Ferrigno}, C., {et~al.} 2013{\natexlab{a}}, The
  Astronomer's Telegram, 4959

\bibitem[{{Papitto} {et~al.}(2013{\natexlab{b}}){Papitto}, {Ferrigno}, {Bozzo},
  {Rea}, {Pavan}, {Burderi}, {Burgay}, {Campana}, {di Salvo}, {Falanga},
  {Filipovi{\'c}}, {Freire}, {Hessels}, {Possenti}, {Ransom}, {Riggio},
  {Romano}, {Sarkissian}, {Stairs}, {Stella}, {Torres}, {Wieringa}, \&
  {Wong}}]{papitto13c}
{Papitto}, A., {Ferrigno}, C., {Bozzo}, E., {et~al.} 2013{\natexlab{b}}, \nat,
  501, 517

\bibitem[{{Papitto} {et~al.}(2014){Papitto}, {Torres}, \& {Li}}]{Papitto14}
{Papitto}, A., {Torres}, D.~F., \& {Li}, J. 2014, \mnras, 438, 2105

\bibitem[{{Patruno}(2010)}]{Patruno10}
{Patruno}, A. 2010, \apj, 722, 909

\bibitem[{{Patruno}(2013)}]{Patruno13}
{Patruno}, A. 2013, The Astronomer's Telegram, 5068

\bibitem[{{Patruno} {et~al.}(2014){Patruno}, {Archibald}, {Hessels},
  {Bogdanov}, {Stappers}, {Bassa}, {Janssen}, {Kaspi}, {Tendulkar}, \&
  {Lyne}}]{patruno14}
{Patruno}, A., {Archibald}, A.~M., {Hessels}, J.~W.~T., {et~al.} 2014, \apjl,
  781, L3

\bibitem[{{Patruno} \& {Watts}(2012)}]{patruno12}
{Patruno}, A. \& {Watts}, A.~L. 2012, ArXiv e-prints

\bibitem[{{Pavan} {et~al.}(2013){Pavan}, {Wong}, {Wieringa}, {Tothill},
  {Filipovic}, {Bozzo}, {Ferrigno}, {Papitto}, \& {Romano}}]{Pavan13}
{Pavan}, L., {Wong}, G., {Wieringa}, M.~H., {et~al.} 2013, The Astronomer's
  Telegram, 4981

\bibitem[{{Poutanen}(2006)}]{Poutanen06}
{Poutanen}, J. 2006, Advances in Space Research, 38, 2697

\bibitem[{{Poutanen} \& {Beloborodov}(2006)}]{pb06}
{Poutanen}, J. \& {Beloborodov}, A.~M. 2006, \mnras, 373, 836

\bibitem[{{Poutanen} \& {Gierli{\'n}ski}(2003)}]{pg03}
{Poutanen}, J. \& {Gierli{\'n}ski}, M. 2003, \mnras, 343, 1301

\bibitem[{{Poutanen} \& {Svensson}(1996)}]{ps96}
{Poutanen}, J. \& {Svensson}, R. 1996, \apj, 470, 249

\bibitem[{{Powell} {et~al.}(2007){Powell}, {Haswell}, \& {Falanga}}]{Powell07}
{Powell}, C.~R., {Haswell}, C.~A., \& {Falanga}, M. 2007, \mnras, 374, 466

\bibitem[{{Psaltis} \& {Lamb}(1999)}]{Psaltis99}
{Psaltis}, D. \& {Lamb}, F.~K. 1999, Astronomical and Astrophysical
  Transactions, 18, 447

\bibitem[{{Radhakrishnan} \& {Srinivasan}(1982)}]{r82}
{Radhakrishnan}, V. \& {Srinivasan}, G. 1982, Current Science, 51, 1096

\bibitem[{{Romano} {et~al.}(2013){Romano}, {Barthelmy}, {Burrows}, {D'Elia},
  {Gehrels}, {Holland}, {Kennea}, {Markwardt}, {Marshall}, \&
  {Page}}]{romano13}
{Romano}, P., {Barthelmy}, S.~D., {Burrows}, D.~N., {et~al.} 2013, The
  Astronomer's Telegram, 4929

\bibitem[{{Sanna} {et~al.}(2017){Sanna}, {Pintore}, {Bozzo}, {Ferrigno},
  {Papitto}, {Riggio}, {Di Salvo}, {Iaria}, {D'A{\`i}}, {Egron}, \&
  {Burderi}}]{Sanna17}
{Sanna}, A., {Pintore}, F., {Bozzo}, E., {et~al.} 2017, \mnras, 466, 2910

\bibitem[{{Serino} {et~al.}(2013){Serino}, {Takagi}, {Negoro}, {Ueno},
  {Tomida}, {Nakahira}, {Kimura}, {Ishikawa}, {Mihara}, {Sugizaki}, {Morii},
  {Yamamoto}, {Sugimoto}, {Matsuoka}, {Kawai}, {Usui}, {Ishikawa}, {Yoshii},
  {Yoshida}, {Sakamoto}, {Nakano}, {Tsunemi}, {Sasaki}, {Nakajima},
  {Fukushima}, {Onodera}, {Suzuki}, {Ueda}, {Shidatsu}, {Kawamuro}, {Tsuboi},
  {Higa}, {Yamauchi}, {Yoshidome}, {Ogawa}, {Yamada}, \& {Yamaoka}}]{Serino13}
{Serino}, M., {Takagi}, T., {Negoro}, H., {et~al.} 2013, The Astronomer's
  Telegram, 4961

\bibitem[{{Shahbaz} {et~al.}(2015){Shahbaz}, {Linares}, {Nevado},
  {Rodr{\'{\i}}guez-Gil}, {Casares}, {Dhillon}, {Marsh}, {Littlefair},
  {Leckngam}, \& {Poshyachinda}}]{Shahbaz15}
{Shahbaz}, T., {Linares}, M., {Nevado}, S.~P., {et~al.} 2015, \mnras, 453, 3461

\bibitem[{{Strohmayer} \& {Bildsten}(2003)}]{Strohmayer03}
{Strohmayer}, T. \& {Bildsten}, L. 2003, ArXiv Astrophysics e-prints

\bibitem[{{Titarchuk} {et~al.}(2002){Titarchuk}, {Cui}, \& {Wood}}]{t02}
{Titarchuk}, L., {Cui}, W., \& {Wood}, K. 2002, \apjl, 576, L49

\bibitem[{{Ubertini} {et~al.}(2003){Ubertini}, {Lebrun}, {Di Cocco}, {Bazzano},
  {Bird}, {Broenstad}, {Goldwurm}, {La Rosa}, {Labanti}, {Laurent}, {Mirabel},
  {Quadrini}, {Ramsey}, {Reglero}, {Sabau}, {Sacco}, {Staubert}, {Vigroux},
  {Weisskopf}, \& {Zdziarski}}]{u03}
{Ubertini}, P., {Lebrun}, F., {Di Cocco}, G., {et~al.} 2003, \aap, 411, L131

\bibitem[{{Watts} \& {Strohmayer}(2006)}]{watts06}
{Watts}, A.~L. \& {Strohmayer}, T.~E. 2006, \mnras, 373, 769

\bibitem[{{Wijnands} \& {van der Klis}(1998)}]{Wijnands98}
{Wijnands}, R. \& {van der Klis}, M. 1998, \nat, 394, 344

\bibitem[{{Wilms} {et~al.}(2000){Wilms}, {Allen}, \& {McCray}}]{wilms00}
{Wilms}, J., {Allen}, A., \& {McCray}, R. 2000, \apj, 542, 914

\bibitem[{{Winkler} {et~al.}(2003){Winkler}, {Courvoisier}, {Di Cocco},
  {Gehrels}, {Gim{\'e}nez}, {Grebenev}, {Hermsen}, {Mas-Hesse}, {Lebrun},
  {Lund}, {Palumbo}, {Paul}, {Roques}, {Schnopper}, {Sch{\"o}nfelder},
  {Sunyaev}, {Teegarden}, {Ubertini}, {Vedrenne}, \& {Dean}}]{w03}
{Winkler}, C., {Courvoisier}, T.~J.-L., {Di Cocco}, G., {et~al.} 2003, \aap,
  411, L1

\bibitem[{{Zdziarski} {et~al.}(1996){Zdziarski}, {Johnson}, \&
  {Magdziarz}}]{Zdziarski96}
{Zdziarski}, A.~A., {Johnson}, W.~N., \& {Magdziarz}, P. 1996, \mnras, 283, 193

\bibitem[{{{\.Z}ycki} {et~al.}(1999){{\.Z}ycki}, {Done}, \& {Smith}}]{Zycki99}
{{\.Z}ycki}, P.~T., {Done}, C., \& {Smith}, D.~A. 1999, \mnras, 309, 561

\end{thebibliography}

\end{document}